 \documentclass[twocolumn,astrosymb]{aastex631}

\usepackage{amsmath,amssymb,amsthm}
\usepackage{mathtools}
\usepackage{subfigure}
\usepackage{graphicx}
\usepackage{color,soul}
\usepackage{longtable}
 
 \hypersetup{linkcolor=red,citecolor=green,filecolor=cyan,urlcolor=magenta}

\begin{document}

\title{Pair Plasma Cascade in Rotating Black Hole Magnetospheres with Split Monopole Flux Model}

\author[0000-0001-9003-0737]{Michael C. Sitarz}
\affiliation{Department of Physics and Astronomy, University of Kansas, Lawrence, KS 66045}
\email{mcsitarz@ku.edu}

\author[0000-0001-5987-2856]{Mikhail V. Medvedev}
\affiliation{Department of Physics and Astronomy, University of Kansas, Lawrence, KS 66045}
\email{medvedev@ku.edu}
\affiliation{Laboratory for Nuclear Science, Massachusetts Institute of Technology, Cambridge, MA 02139}
\affiliation{Institute for Advanced Study, School for Natural Sciences, Princeton, NJ 08540}
\affiliation{Department of Astrophysical Sciences, Princeton University, Princeton, NJ 08544}

\author[0000-0001-6805-9787]{Alexander L. Ford}
\altaffiliation{Adjunct Research Support}
\affiliation{Department of Physics and Astronomy, University of Kansas, Lawrence, KS 66045}
\affiliation{Northrop Grumman Corporation, Boulder, CO 80301}
\email{alexlford@ku.edu}



\begin{abstract}

An electron-positron cascade in the magnetospheres of Kerr Black Holes (BH) is a fundamental ingredient to fueling the relativistic $\gamma$-ray jets seen at the polar regions of galactic supermassive BHs (SMBH). This leptonic cascade occurs in the ``spark gap'' region of a BH magnetosphere where the unscreen electric field parallel to the magnetic field is present, hence it is affected by the magnetic field structure. A previous study explored the case of a thin accretion disk, representative of Active Galactic Nuclei (AGN). Here we explore the case of a quasi-spherical gas distribution, as is expected to be present around the SMBH Sgr A* in the center of our Milky Way galaxy, for example. The properties and efficiency of the leptonic cascade are studied. The findings of our study and the implications for SMBH systems in various spectral and accretion states are discussed. The relationships and scalings derived from varying the mass of the BH and background photon spectra are further used to analyze the leptonic cascade process to power jets seen in astronomical observations. In particular, one finds the efficiency of the cascade in a quasi-spherical gas distribution peaks at the jet axis. Observationally, this should lead to a more prominent jet core, in contrast to the thin disk accretion case, where it peaks around the jet-disk interface. One also finds the spectrum of the background photons to play a key role. The cascade efficiency is maximum for a spectral index of two, while harder and softer spectra lead to a less efficient cascade.

\end{abstract}



\section{Introduction} \label{sec:intro}

\indent The environment surrounding rotating supermassive black holes (SMBH) are not only fully general relativistic but can contain an extreme astrophysical plasma environment. These systems are surrounded by a plasma-rich magnetosphere and emit highly energetic $\gamma$-ray jets emanating from near the horizon. Blandford and Znajek (BZ) \citep{1} were the first to theorize how the magnetospheric plasma could power the $\gamma$-ray jets by converting the rotational energy to electromagnetic energy via the Poynting vector ($\Vec{S}$). This process can be expressed using the spin down luminosity \citep{2}
\begin{equation}\label{Eq:spin_down_lum}
    L_{sd} = \Omega_F(\omega_H - \Omega_F)B_{\perp}^2r_H^4c^{-1},
\end{equation}
where $\Omega_F$ denotes the angular frequency of the magnetic field lines rotating about the BH, $\omega_H = ac/r_sr_H$ defines the angular frequency, $r_s = 2GMc^{-2}$ is the Schwarzschild radius, $r_H$ is the horizon radius $r_H = GM/c^2 + [(r_s/2)^2 - a^2]^{1/2}$, $a = J/Mc$ is the spin parameter of a BH, and $B_{\perp}$ is the perpendicular component of the magnetic field with respect to the BH surface. \\

\indent \citep{3}, \citep{4}, \citep{5}, and others \citep{6,7,8,9} further proposed a mechanism to fill the magnetosphere of a rotating, uncharged black hole with plasma that could then undergo the BZ mechanism. A similar mechanism works in a pulsar magnetosphere, see \citep{10,11,12,13,14,15,16,17,18,19} and the references therein. These mechanisms for powering jets often assume that the magnetospheric plasma is force free:
\begin{equation}\label{Eq:mag_force_free}
    \rho_e\Vec{E} + \Vec{j}/c \times \Vec{B} = 0,
\end{equation}
where $\rho_e$, $\vec{E}$, $\vec{j}$, $c$, and $\vec{B}$ are the charge density, electric field, current density, speed of light, and magnetic field (respectively). For a force free plasma, the magnetic pressure ($P_B = B^2/2\mu_0$) must also greatly 
exceed the plasma pressure ($P_P = nk_BT$). This allows nonmagnetic forces to be neglected. The region where $e^\pm$ plasma can be produced is called the ``spark gap'' or just the ``gap.'' \\

\indent If the conditions are right in the gap, an $e^{\pm}$ plasma cascade occurs, filling the force free magnetosphere with plasma which in turn powers the $\gamma$-ray jets through the BZ mechanism. This cascade exists in a cyclical lifespan described in the following paragraph.\\

\indent Assume the Kerr BH (KBH) is effectively in a plasma with a magnetic field that threads the horizon. Also assume that there is a bath of soft photons originating from the accretion disk surrounding the KBH. The KBH now can act as a conductor, the spin of the BH creates a motional electric field with frame dragging effects. If the component of the electric field parallel to the magnetic field is non-zero then particle acceleration can occur. Electrons and positrons accelerated by this parallel electric field have a chance to Compton up-scatter the soft photons existing around a BH (produced by an accretion flow, for example). The up-scattered $\gamma$-ray photons have a chance to interact with each other and produce and electron-positron pair: 
\begin{equation}\label{Eq:pair_production}
    \gamma + \gamma \rightarrow e^+ + e^-,
\end{equation}
provided their total energy in the center-of-mass frame exceeds $2m_2c^2$. With two new charged particles, this process of $e^\pm$ acceleration, Compton up-scattering, and $\gamma\gamma$ pair production repeats. \\

\indent This process has two threshold parameters that can dictate the processes efficiency and productivity. The size of the region of plasma (referred to now as the gap) must be large enough. The larger the gap, the more region is available for a particle to accelerate and  scatter, the higher the probability of scattering. Second, the photons must have enough energy to exceed the rest mass of the secondary particles. If the mass-energy threshold is not met, the pair production will not occur.\\

\indent This work will extend the study of \citep{20} by sweeping the parameter space of BH mass and spectral index using a different magnetic flux function than that of previous works:
\begin{equation}\label{Eq:new_flux}
    \Psi(\theta) = (1 - cos(\theta)),
\end{equation}
which represents the force-free ``split-monopole'' configuration. This new model more accurately reenacts what occurs in nature which can further our understanding of various systems seen in observations without the use of assumptions or limiting the types of environments covered by the previous flux model. Section $2$ will introduce the mathematical framework of the study is conducted in along with the cascade mechanism expressions and system of equations that are solved along with the boundary conditions imposed on the system \citep{7,21}. \S3 details the techniques involved in solving the rigid system of equations that govern the BL processes. The data of the study is presented in \S4, followed by a results discussion and conclusion in \S5.

\section{Mathematical System}
\subsection{Metric and Coordinate System}
\indent Given a KBH with mass $M$, angular momentum $J$, and charge $Q $, we assume the force free magnetosphere surrounding the surface is stationary and asymmetric. To study this system, we employ the Kerr metric \citep{22} expressed through the Boyer-Lindquist coordinate system \citep{23}. This system of coordinates describes space-time in the vicinity of a BH. The coordinate transform for an oblate spheroid is as follows:
\begin{subequations}
    \begin{equation}\label{Eq:x_coord}
    x = \sqrt{r^2 + a^2}sin(\theta)cos(\phi),
    \end{equation}
    \begin{equation}\label{Eq:y_coord}
        y = \sqrt{r^2 + a^2}sin(\phi)sin(\phi),
    \end{equation}
    \begin{equation}\label{Eq:z_coord}
        z = rcos(\theta),
    \end{equation}
\end{subequations}
with the time dimension still expressed by $t$. \\

\indent In the Kerr Black Hole scenario, charge $Q = 0$. This constraint simplifies the Boyer-Lindquist coordinates while also allowing for a dimensionless spin parameter, $a$, to be defined. Black hole characteristics angular momentum $J$ and charge $Q$ are constrained by mass $M$, which may take any positive value. The constraint equation
\begin{equation}\label{Eq:BH_Constraint}
    \frac{Q^2}{GM^2} + \frac{J^2c^2}{GM^2} \leq J,
\end{equation}
simplifies to
\begin{equation}\label{Eq:Ang_Mom_Relation}
    J \leq \frac{GM}{c^2}.
\end{equation}
This can be rearranged to express limits on the value
\begin{equation}\label{Eq:Kerr_Relation}
    0 \leq \frac{Jc}{GM^2} \leq 1.
\end{equation}
In natural units ($c = G = 1$), $J/M$ is labeled the Kerr Parameter. This work will employ a unit system in which the spin parameter $a \equiv J/Mc$ (found by multiplying the original expression by $GMc^{-2}$). With $a$ defined, the horizon radius can be re-expressed using the gravitation radius $r_g = GM/c^2$ and spin parameter, $r_H = r_g + \sqrt{r_g^2 - a^2}$. With these parameters, the Kerr metric under the Boyer-Lindquist coordinates system can be properly expressed using coordinates $(t, r, \theta, \psi)$.\\

\indent The Kerr metric can be succinctly expressed via tensor notation 
\begin{equation}\label{Eq:Kerr_metric_line}
    ds^2 \equiv g_{tt}dt^2 + 2g_{t\psi}dtd\psi + g_{\psi\psi}d\psi^2 + g_{rr}dr^2 + g_{\theta\theta}d\theta^2.
\end{equation}
Here we can note that the term $dtd\psi$ implies a time-space coupling in the plane of rotation as long as $J > 0$ holds. The field tensors can be defined in terms of length scales $\Delta$ (called the discriminant), $\Sigma$, $A$, $r_H$, and $a$:
\begin{subequations}
    \begin{equation}\label{Eq:gtt}
        g_{tt} \equiv -\left(\frac{r^2 - r_sr + a^2 - a^2sin^2(\theta)}{r^2 + a^2cos^2(\theta)}\right)c^2,
    \end{equation}
    \begin{equation}\label{Eq:gtp}
        g_{t\psi} \equiv \frac{-r_srsin^2(\theta)ac}{r^2 + a^2cos^2(\theta)},
    \end{equation}
    \begin{equation}\label{Eq:gpp}
        g_{\psi\psi} \equiv \frac{(r^2 + a^2)^2 - (r^2 - r_sr + a^2) a^2 sin^4(\theta)}{r^2 + a^2cos^2(\theta)},
    \end{equation}
    \begin{equation}\label{Eq:grr}
        g_{rr} \equiv \frac{r^2 + a^2cos^2(\theta)}{ r^2 - r_sr + a^2},
    \end{equation}
    \begin{equation}\label{Eq:gthth}
        g_{\theta\theta} \equiv r^2 + a^2cos^2(\theta).
    \end{equation}
\end{subequations}

\indent Using the above expressions, the line element $ds^2$ can be fully expressed in terms of the Boyer-Lindquist coordinates and BH base characteristics (shown in full in Appendix A). From the full expression, it can be condensed using the spin parameter, radius $r_s$, length scales $\rho^2$, $\Delta$, $\Sigma^2$, and scalar functions introduced by \citep{24} $\alpha$, $\omega$, and $\Bar{\omega}$. These new condensing expression are defined as:
\begin{subequations}
    \begin{equation}\label{Eq:rho2}
        \rho^2 \equiv r^2 + a^2cos^2(\theta),
    \end{equation}
    \begin{equation}\label{Eq:Delta}
        \Delta \equiv r^2 + a^2 - r_s,
    \end{equation}
    \begin{equation}\label{Eq:Sigma2}
        \Sigma^2 \equiv (r^2 + a^2) - a^2\Delta sin^2(\theta),
    \end{equation}
    \begin{equation}\label{Eq:alpha}
        \alpha \equiv \rho\Sigma^{-1}\Delta^{1/2},
    \end{equation}
    \begin{equation}\label{Eq:omega}
        \omega \equiv acr_s\Sigma^{-2},
    \end{equation}
    \begin{equation}\label{Eq:omega_bar}
        \Bar{\omega} \equiv \Sigma\rho^{-1}sin(\theta).
    \end{equation}
\end{subequations}
The line element can now be expressed in its final form
\begin{multline}\label{Eq:Kerr_metric_final_form}
    ds^2 = (\omega^2\Bar{\omega}^2 - \alpha^2)dt^2 - 2\omega\Bar{\omega}^2dtd\psi \\ + \rho^2\Delta^{-1}dr^2 + \rho^2d\theta + \Bar{\omega}d\psi^2.
\end{multline}

\subsection{Black Hole Magnetosphere Field Equations}
\indent To describe the electromagnetic fields and phenomena that occur in the magnetosphere of a KBH, we apply the $3 + 1$ split rule of electrodynamics laws \citep{24}; where $4D$ space-time is split into global time $t$ and $3D$ curved space.\\

\indent Before moving on, the fields inside the gap must be defined according to the force free condition on the gap. We recall the force free condition for the magnetosphere ($\rho_e\vec{E} + j/c \times \vec{B} = 0$) where we now define
\begin{equation}\label{Eq:charge_density_particles}
    \rho_e = e(n^+ - n^-),
\end{equation}
as the charge density in the plasma contributed by the positrons and electrons. This then recovers the starting Poisson equation
\begin{equation}\label{Eq:starting_poisson}
    \nabla \cdot \vec{E} = 4\pi\rho_e.
\end{equation}
For a force free magnetosphere that follows the degenerate condition $\vec{E}\cdot\vec{B} = 0$, see \citep{25}, the toroidal magnetic field can be written as 
\begin{equation}\label{Eq:toroidal_mag}
    \vec{B}_T = \frac{-2I(\Psi)}{\alpha\Bar{\omega}}\hat{\phi},
\end{equation}
where $I(\Psi)$ is the current flowing back towards the BH surface with constant flux $\Psi(r, \theta)$ (the actual form of the flux function is discussed below). The poloidal field is defined as. 
\begin{equation}\label{Eq:poloidal_mag}
    \vec{B}_P = \frac{\nabla\Psi\hat{\phi}}{2\pi\Bar{\omega}}.
\end{equation}
From this point, we use the notation for fields $\vec{F}_T = \vec{F}_\perp$ and $\vec{F}_P = \vec{F}_\parallel$, interchangeably in relation to the surface of the KBH. To find the electric field components, we recall the gap has a frozen-in condition. The electric field can either be calculated through
\begin{equation}\label{Eq:parallel_elec}
    \vec{E}_\parallel = \left(\frac{\vec{B}}{B}\right) \cdot \vec{E} = \left(\frac{\vec{B}}{B}\right) \cdot (-\nabla\Psi),
\end{equation}
or more simply
\begin{equation}\label{Eq:poloidal_elec}
    \vec{E}_P = \frac{-\vec{v}_F}{c} \times \vec{B}_P.
\end{equation}
Here we define the rotational velocity of the field lines measured by a ZAMO (zero angular momentum observer) by
\begin{equation}\label{Eq:field_line_velo}
    \vec{v}_F = \frac{(\Omega_F - \omega)\Bar{\omega}}{\alpha}\hat{\phi},
\end{equation}
where the angular velocity of the field lines $\Omega_F = d\phi/dt$ and angular velocity of BH $\omega = (d\tau/dt)_{ZAMO}$. From these prescriptions, the electric field components recovered are 
\begin{equation}\label{Eq:toroidal_elec}
    \vec{E}_T = 0,
\end{equation}
and 
\begin{equation}\label{Eq:Poloidal_Electric_Field}
    \vec{E}_P = - \frac{\Omega_F - \omega}{2\pi\omega}\nabla\Psi.
\end{equation}
This gives a Poisson equation of the degenerate, force-free magnetosphere as
\begin{equation}\label{Eq:GJ_charge_density}
    \rho_{GJ} = \frac{\nabla \cdot \vec{E}_P}{4\pi}.
\end{equation}
This is defined as as the Goldreich-Julian charge density \citep{26}. Rewriting the electric field as 
\begin{equation}\label{Eq:GJ_electric_field}
    \vec{E} = \vec{E'} - \vec{E}_P,
\end{equation}
the Poisson equation inside the gap is finally 
\begin{equation}\label{Eq:gap_Poisson}
    \nabla \cdot \vec{E}_P = 4\pi(\rho_e - \rho_{GJ}).
\end{equation}
For a tensor notation based discussion of $\rho_{GJ}$ please see \citep{21}.\\

\indent The last expression to be defined is the flux $\Psi(r, \theta)$. Previous works used a double split monopole defined by
\begin{equation}\label{Eq:old_flux_expr}
    \Psi(\theta) = \Psi_Msin^2(\theta).
\end{equation}
This flux function was used to mime the existence of a thin accretion disk in the vicinity of a horizon. In the current work we instead use the split-monopole configuration. This configuration naturally occurs in numerical simulations when the accreting gas is hot. This corresponds to a geometrical thick accretion flow, such as advection or convection dominated accretion flow (ADAF and CDAF). Such accretion flows are generally x-ray bright (in contrast to thin disk accretion), so they are likely present in many AGNs. Splitting the monopole into two hemispherical planes \citep{27} can be expressed as 
\begin{equation}\label{Eq:split_monopole}
    \Psi(\theta) = 
    \begin{cases}
    \Psi_0(1 - cos(\theta)),& \theta \ \epsilon \ [0, \pi/2)\\
    \Psi_0(1 + cos(\theta),& \theta \ \epsilon \ (\pi/2, \pi].
\end{cases}
\end{equation}
As one can see, a discontinuity arises at the equatorial plane ($\theta = \pi/2$) where a current sheet is present. For this work, $\theta$ will be swept from $0$ to $\pi/2$, avoiding the discontinuity and assuming symmetry in the magnetosphere.\\

\indent The field configuration derived above is chosen for the following reasons. First, this depicts a simplified model of an accreting BH where the jets are located in the polar regions and the disk is associated with the equatorial plane. Second, inside the gap, the field will always be radial. Therefore, the gap and the plasma production within it is nicely captured by the above configuration. \\

\indent While the gap in the magnetosphere has been mentioned and the fields inside defined, the null surface has yet to be fully explained. There is a surface that exists in the magnetosphere where $\rho_{GJ} = 0$. This null area has very strong $\vec{E}_{||}$ parallel to the magnetic field lines. There exists a surface of charge deficit around this electric field. As described above, an $e^\pm$ cascade in this region is required to maintain the force free condition  on the gap that will power the BH jets. 

\subsection{$e^\pm$ Cascade Process}
\begin{figure*}
    \centering
    \includegraphics[scale=0.60]{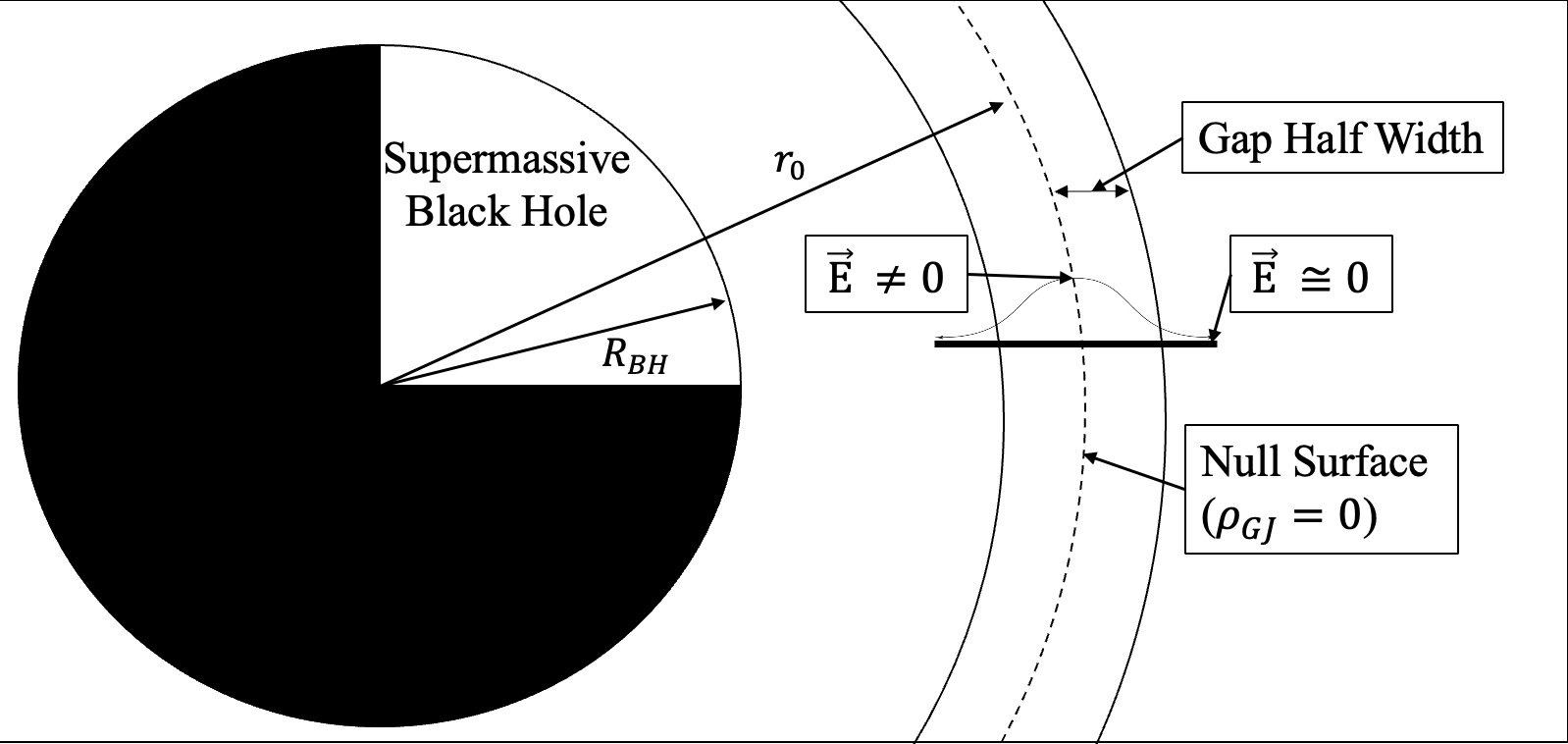}
    \caption{Schematic depiction of the spark gap and SMBH system. Generally $H << r_0$, unless the cascade is not efficient of present at all.}
    \label{Fig:Gap_Schematic}
\end{figure*}
\indent With the global field equations established, we can shift focus to the gap itself and the equations that govern the cascade system. Recall that $\vec{E}$ can no longer be screened out near the gap due to insufficient plasma, giving rise to to the important $\vec{E}_{||}$ field. This phenomena generates the flowing, charge separated plasma seen in the magnetosphere. A schematic of the system is seen in (Fig. \ref{Fig:Gap_Schematic}). While the poloidal magnetic field, $\vec{B}_p$, (Eq. \ref{Eq:poloidal_mag}) may be oblique in general, it can be constrained to the full perpendicular case for this discussion. In this circumstance, the acceleration of charges by the parallel electric field (Eq. \ref{Eq:Poloidal_Electric_Field}) is at its most efficient and therefore the cascade system itself is at its most efficient. Further limiting the system into one dimension, define the $x$ coordinate to be perpendicular to the ''null surface`` and zero at the center of the gap.
\begin{equation}\label{Eq:gap_center_coord}
    x = r - r_0,
\end{equation}
with $r_0$ being the null surface of the gap. For a spherical gap, with x increasing outwardly, $r_0$ is the radius at which $\rho_{GJ}$ is zero. With the dimensional reduction, the Poisson equation (Eq. \ref{Eq:gap_Poisson}) now reads
\begin{equation}\label{Eq:Poisson_dimen_reduc}
    \frac{dE_{||}}{dx} = 4\pi[e(n^+ - n^-) - \rho_{GJ}].
\end{equation}
With the gap ($r_0$) considerably smaller than $H$ (less than $1\%$ of $r_H$), the Poisson equation can be adjusted by expanding $\rho_{GJ}(x,\theta)$ about $x = 0$ 
\begin{equation}\label{Eq:Poisson_reduc_GJ}
    \frac{dE_{||}}{dx} = 4\pi[e(n^+ - n^-) - A_{\theta}x].
\end{equation}
This recovers the expansion coefficient $A_{\theta} = \partial_r\rho_{GJ}(x=0,\theta)$, which is on the order of ($\Omega_F B/2\pi ce$).\\

\indent Inside the gap, $E_{||}$ is accelerating the $e^{\pm}$ charges and Compton scattering them off of UV-photons from the accretion disk, making the charges rapidly loose momentum. Though, longitudinal motion is still steady due to $E_{||}$. From this set up, the equation of motion for a charge can be written using the Lorentz factor $\Gamma$, the Thompson scattering cross section $\sigma_T$, and the energy density of the background radiation field $U_b$;
\begin{equation}\label{Eq:equation_of_mot}
    m_ec^2\frac{d\Gamma}{dx} = eE_{||} - (\Gamma^2 - 1)\sigma_TU_b.
\end{equation}
The Thompson scattering cross section is defined by 
\begin{equation}\label{Eq:Thompson_Scatter}
    \sigma_T = \frac{8\pi}{3}\left(\frac{q^2}{4\pi\epsilon_0mc^2}\right)^2 = \frac{8\pi\alpha^2\hbar^2c^2}{3m_e^2c^4}.
\end{equation}
The second term in the equation of motion (Eq. \ref{Eq:equation_of_mot}) represents Compton drag, which may be overestimated in a general sense due to simplification. This simplification still holds for pair production. At the boundaries of the gap, $E_{||}$ is expected to be zero. Except for at this condition (everywhere else $E_{||} > 0$), the RHS of (Eq. \ref{Eq:equation_of_mot}) cancels out in the first order where $\Gamma$ reaches a terminal value
\begin{equation}\label{Eq:Gamma_terminal_val}
    \Gamma(x) = \sqrt{1 + \frac{eE_{||}(x)}{\sigma_TU_b}}.
\end{equation}

\indent The newly created $e^{\pm}$ have perpendicular momentum $P_{\perp}$. This momentum is gradually lost as the charges scatter off background radiation photons in a drag length defined by 
\begin{equation}\label{Eq:drag_length}
    l_{drag} = \frac{\Gamma m_e c^2}{\Gamma^2 \sigma_TU_b} = \frac{m_e c^2}{\sqrt{eE_{||}(x)\sigma_TU_b}}.
\end{equation}
This drag length is smaller than the Compton mean free length\footnote{The mean free length is the inverse of the cross section and average length between collisions. For Compton scattering, $l_{drag}$ on the order of $10^{10}$ cm.} where the charges scatter off background soft photons to produce the desired $\gamma$-ray photons required for pair production. Using these lengths, and the length of the half gap, $H$, the following relation is defined
\begin{equation}\label{Eq:length_relation}
    l_{drag} < l_c < H.
\end{equation}
From this relation, two major approximations are deduced: charges migrate in one direction and the charges have a monoenergetic motion with respect to $\Gamma(x)$. This relation holds due to X-ray and UV photons contributing to the drag term. Despite the various non-simplified drag terms, the charges can still produce the necessary $\gamma$-ray photons with energies up to $\Gamma m_ec^2$. These $\gamma$-ray photons are emitted from inverse Compton scattering \citep{6}, not curvature radiation or synchrotron radiation. These new $\gamma$-ray photons are free to pair produce charges by collisions with background photons.\\

\indent For a $\gamma$-ray photon of energy $\epsilon_\gamma m_e c^2$ to undergo pair production it must collide with a background photon with energy $\epsilon_s m_e c^2$. The photons must satisfy
\begin{equation}\label{Eq:photon_energy_relat}
    \epsilon_s\epsilon_\gamma \geq \frac{1}{1 - \mu},
\end{equation}
where $\mu$ is the cosine of the collision angle $\theta$. The minimum energy required for production occurs when the photons collide head on ($\mu = -1$) with the most energetic soft photon collision occurring with $\epsilon_s = \epsilon_{max}$. Only $\gamma$-rays at energies greater than the above minimum energy ($\epsilon_{max}^{-1}$) can contribute to pair production.

\subsection{Basic Physical System Equations}
\indent With the physical process of the cascade described, the equations governing the charges and fields can now be examined in the cascade context. To begin, it is necessary and helpful to define the motion of the electrons and positrons in the gap. The motion of the charges is set by the direction of the current as it moves towards the polar regions. For this system, the electrons will migrate inward toward the BH while the positrons migrate outward away from the BH. This set-up is accurate in lower latitudes (small distance away from BH) and appears opposite in high latitudes. But, the treatment of physics holds in each scenario as the signs of the various terms in (Eq. \ref{Eq:Poisson_dimen_reduc}) resolve themselves and create the same effect. This research will consider low latitudes only. The speed of these mobile charges can be defined as 
\begin{equation}\label{Eq:particle_velocity}
    \vec{v}_{e^\pm} = c\beta(x),
\end{equation}
with 
\begin{equation}\label{Eq:beta_factor}
    \beta(x) = \sqrt{1 - \frac{1}{\Gamma^2(x)}},
\end{equation}
defined for brevity. The speed is shared by all charges due to the monoenergetic assumption. From this, the continuity equation \citep{28} then can be defined as 
\begin{multline}\label{Eq:continuity_equation}
    \pm\frac{d}{dx}[n^\pm(x)\beta(x)] = \\ \int_0^\infty \eta_p(\epsilon_\gamma)[F^+(x, \epsilon_\gamma) + F^-(x,\epsilon_\gamma)]d\epsilon_\gamma.
\end{multline}
$F^\pm$ are the Boltzmann equations for the $\gamma$-ray photons, defined later in the section (Eq. \ref{Eq:photon_boltzmann}). $\eta_p(x)$ is defined as the angle averaged pair production redistribution function expressed via 
\begin{equation}\label{Eq:pair_redist_factor}
    \eta_p(x) = \frac{1}{2}\int_{-1}^1 d\mu \int_{\frac{1}{(1-\mu)\epsilon_\gamma}}^{\epsilon_{max}}d\epsilon_s\sigma_p\frac{dN_s}{d\epsilon_s}.
\end{equation}
Here, $\sigma_p$ representing the cross section of pair production in collisions between photons of energies $m_ec^2\epsilon_s$ and $m_ec^2\epsilon_\gamma$ moving angle $\mu$ with respect to each other;
\begin{multline}\label{Eq:pair_cross_section}
    \sigma_p \equiv \frac{3}{16}\sigma_T(1-\nu^2) \\ \times
    \left[(3-\nu^4)\ln\left(\frac{1+\nu}{1-\nu}\right) - 2\nu(2-\nu^2)\right],
\end{multline}
only when the energy relation is met (Eq. \ref{Eq:photon_energy_relat}) does the above expression have a non-vanishing value. With the $\nu$ being defined as,
\begin{equation}\label{Eq:nu_factor}
    \nu(\mu,\epsilon_\gamma,\epsilon_s) \equiv \sqrt{1 - \frac{2}{1-\mu}- \frac{1}{\epsilon_s\epsilon_\gamma}}.
\end{equation}
$n^\pm(x)$ and $F^\pm(x,\epsilon_\gamma)$ represent the number density of outwardly/inwardly moving particles and $\gamma$-ray photons in the $\pm$ x-direction in a non-dimensional energy interval $\epsilon_\gamma \approx \epsilon_\gamma + d\epsilon_\gamma$, respectively. Finally, $dN_s/d\epsilon_s$ is the number density of soft background photons in a non-dimensional energy interval $\epsilon_s \approx \epsilon_s + d\epsilon_s$. The $\gamma$-ray photons created by inverse Compton scattering are highly beamed in the same direction as the charges, $e^\pm$, which is to say that the distribution functions can be fully described by $F^\pm$.\\

\indent Current is conserved as it is carried along the field lines. From one combination of the continuity equation (Eq. \ref{Eq:continuity_equation})
\begin{equation}\label{Eq:cont_combo_one}
    \frac{d}{dx}\left[(n^+(x) + n^-(x))\beta(x)\right] = 0,
\end{equation}
which recovers
\begin{equation}\label{Eq:current_density}
    \left[(n^+(x) + n^-(x))\beta(x)\right] = \frac{j_0}{e}.
\end{equation}
$E_{||}$ must go to zero at the borders of the gap, which occurs when $j_0 = j_{crit}$. $j_0$ is now defined as the current density along a field line, out-flowing from the gap at a constant rate. When $j_0$ takes the value, within order of magnitude \citep{24}, of $j_0 \approx 10^{-15} \left(\frac{a}{M}\right)\left(\frac{M}{10^8 M_\odot}\right)\left(\frac{B}{10^4 G}\right)\left(\frac{abamp}{cm^2}\right)$ then energy and angular momentum can be effectively extracted from the KBH. Another combination of the continuity equation recovers
\begin{multline}\label{Eq:cont_combo_two}
    \frac{d}{dx}\left([n^+(x) - n^-(x)]\beta(x)\right) = \\ 2\int_0^\infty\eta_p(\epsilon_\gamma)[F^+(x,\epsilon_\gamma) + F^-(x,\epsilon_\gamma)]d\epsilon_\gamma.
\end{multline}
In place of the original continuity equation (Eq. \ref{Eq:continuity_equation}), the two combination of the continuity equation presented above (Eq. \ref{Eq:current_density} and Eq. \ref{Eq:cont_combo_two}) will be used.\\

\indent The single dimensional motion of $\gamma$-ray photons obeys a Boltzmann equation of the form
\begin{equation}\label{Eq:photon_boltzmann}
    \pm\frac{\partial}{\partial x}F^\pm(x, \epsilon_\gamma) = \eta_cn^\pm \beta(x) - \eta_pF^\pm.
\end{equation}
Above, $\eta_c(x)$ is the Compton redistribution function
\begin{equation}\label{Eq:compton_redist}
    \eta_c(\epsilon_\gamma, \Gamma) \equiv \int_{\epsilon_{min}}^{\epsilon_{max}} d\epsilon_s\frac{dN_s}{d\epsilon_s}\sigma_{KN}(\epsilon_s\Gamma)\delta(\epsilon_s - \Gamma^2\epsilon_\gamma),
\end{equation}
with $\sigma_{KN}$ as the Klein-Nishina \citep{29} cross section defined by \citep{30}
\begin{multline}\label{Eq:klein_nishina}
    \sigma_{KN}(z) \equiv \frac{3}{4}\sigma_T \left[ \frac{1 + z}{z^3}\left(\frac{2z(1+z)}{1 + 2z} - \ln(1 + 2z) \right) + \right. \\ \left. \frac{\ln(1 + 2z)}{2z} - \frac{1 + 3z}{(1 + 2z)^2} \right].
\end{multline}
There is the assumption that energy transfer from $e^\pm$ with Lorentz factor $\Gamma$ to a photon with incident energy $m_e c^2 \epsilon_s$ is approximately $ m_e c^2\Gamma^2\epsilon_s$ in the Compton redistribution function. This holds in its simplified form, but can be more complex if need be.\\

\indent The migrating $e^\pm$'s and $\gamma$-rays are described by the differential equations (Eqs. \ref{Eq:Poisson_reduc_GJ}, \ref{Eq:equation_of_mot}, \ref{Eq:cont_combo_two}, \ref{Eq:photon_boltzmann}) while the number density functions $n^\pm$ are related by (Eq. \ref{Eq:current_density} and Eq. \ref{Eq:photon_boltzmann}) (independent of each other). This sums to a system of five differential equations to be solved. \\

\indent For the system of five equations to be solved, there must first be assumptions about the background radiation field made first. The spectral number density of background radiation per unit interval of $\epsilon_s$ can be modeled by a power law
\begin{equation}\label{Eq:spec_num_dens}
    \frac{dN_s}{d\epsilon_s} = C(\alpha)\epsilon_s^{-\alpha}.
\end{equation}
This $C(\alpha)$ term is a decreasing function with respect to $\alpha$ expressed via
\begin{equation}\label{Eq:c_alpha}
    C(\alpha)  \equiv \frac{2 - \alpha}{\epsilon_{max}^{2-\alpha} - \epsilon_{min}^{2-\alpha}}\frac{U_b}{m_ec^2},
\end{equation}
with epsilon terms defining the cutoff of the spectrum. $U_b$ is the background radiation field's energy density.  

\subsection{Non-Gray Analysis of the $\gamma$-Ray Distribution}
\indent Examining the expression for $\eta_p$ (Eq. \ref{Eq:pair_redist_factor}), no gray approximation that can be attributed to it to solve the Boltzmann equation (Eq. \ref{Eq:photon_boltzmann}) due to its $\epsilon_\gamma$ dependence. To rectify this, $\epsilon_\gamma$ is split into bins so that $\eta_p$ is approximated in each bin. To do this, let $\xi_i$ and $\xi_{i - 1}$ be the upper and lower limits of the $i^{th}$ bin (limits are sufficiently close). Using this, the right hand side  of (Eq. \ref{Eq:cont_combo_two}) using a summation instead of solving the following integral types
\begin{equation}\label{Eq:photon_integral}
    \int_{\xi_{i-1}}^{\xi_i}\eta_p(\epsilon_\gamma)F^\pm(x,\epsilon_\gamma)d\epsilon_\gamma \approx \eta_{p,i}f_i^\pm(x),
\end{equation}
where
\begin{equation}\label{Eq:eta_photon}
    \eta_{p,i} \equiv \eta_p\left(\frac{\xi_{i-1} - \xi_i}{2}\right),
\end{equation}
and
\begin{equation}\label{Eq:small_f_photon}
    f_i^\pm = \int_{\xi_{i-1}}^{\xi_i} F^\pm(zx,\epsilon_\gamma)d\epsilon_\gamma.
\end{equation}
In place of integral (\ref{Eq:cont_combo_two}) the following is used
\begin{equation}\label{Eq:energy_sum_relat}
    \frac{d}{dx}\left[(n^+(x) - n^-(x))\beta(x)\right] = 2\Sigma^\chi_{i=1} \eta_{p,i}[f_i^+(x) + f_i^-(x)],
\end{equation}
with $\chi$ number of energy bins. Then (Eq. \ref{Eq:photon_boltzmann}) can be integrated  over an energy bin between each limit to form
\begin{equation}\label{Eq:derivative_f_photon}
    \pm\frac{d}{dx}f^\pm_i(x) = \eta_{c,i}(\Gamma)n^\pm(x)\beta(x) - \eta_{p,i}f^\pm_i(x),
\end{equation}
where
\begin{equation}\label{Eq:eta_compton}
    \eta_{c,i}(\Gamma) = \int_{\xi_{i-1}}^{\xi_i}\eta_c(\epsilon_\gamma, \Gamma)d\epsilon_\gamma.
\end{equation}
The system of equations now composes of (Eqs. \ref{Eq:Poisson_reduc_GJ}, \ref{Eq:equation_of_mot}, \ref{Eq:energy_sum_relat}, \ref{Eq:derivative_f_photon}).

\subsection{Gap Boundary Conditions}
\indent To fully study the physical processes and environment within the gap, it is helpful to consider the case(s) when the functions of $E_{||}, \Gamma, n^+$ and $f_i^\pm$ have symmetric conditions. These boundary conditions are defined in a way so symmetries allow the conditions to be set at the center of the gap ($x = H$) and at the edge of the gap ($x = 0$), allowing only integration over half of the gap. This gap width should only be within a few percents of the black hole radius. The gap boundary is defined when the parallel electric field vanishes. Using (Eq. \ref{Eq:Poisson_dimen_reduc}) at $x = H$ and $E_{||} = 0$, a smooth curve resulting in $\frac{dE_{||}}{dx} = 0$ at the half gap width can be found with (Eq. \ref{Eq:current_density}) and 
\begin{equation}\label{Eq:current_density_gap}
    j_0\beta(x) - A_\theta x = 0.
\end{equation}

\indent To begin, $E_{||}$ should not change sign within the gap and the function should vanish at the boundaries of the gap itself. With $H << r_{0}$ (Eq. \ref{Eq:gap_center_coord}), we may assume that $E_{||}$ is an even function with respect to $x$. This condition also applies to $\Gamma$;
\begin{equation}\label{Eq:symmetry_E_Gamma}
    E_{||}(x) = E_{||}(-x); \ \Gamma(x) = \Gamma(-x).
\end{equation}
Next, the assumption of functional symmetry is imposed on $n^\pm$ (particles)
\begin{equation}\label{Eq:boundary_part_flux}
    n^+(x) = n^-(-x),
\end{equation}
and $F^\pm$ ($\gamma$-ray photons)
\begin{equation}\label{Eq:boundary_photon}
    F^+(x, \epsilon_\gamma) = F^-(-x, \epsilon_\gamma).
\end{equation}
The consequences of the symmetric functions include essential requirements and features of the pair production cascade. Therefore, (Eqs. \ref{Eq:Poisson_reduc_GJ}, \ref{Eq:equation_of_mot}, \ref{Eq:energy_sum_relat}, \ref{Eq:derivative_f_photon}) are solved within $0 \leq x \leq H$.\\

\indent With these functional conditions imposed the derivation of the full boundary conditions at $x = 0$ and $x = H$ can begin. First, we derive the conditions for the inner boundary. From (Eqs. \ref{Eq:Poisson_reduc_GJ}, \ref{Eq:equation_of_mot}, \ref{Eq:symmetry_E_Gamma})
\begin{equation}\label{Eq:in_bound_gam}
    \frac{d\Gamma}{dx} = 0,
\end{equation}
which is equivalent to
\begin{equation}\label{Eq:in_E_par}
    E_{||} = \frac{\sigma_TU_b\Gamma^2}{e},
\end{equation}
is found for $x = 0$. From (Eqs. \ref{Eq:current_density} and \ref{Eq:boundary_part_flux}), $x = 0$ also recovers
\begin{equation}\label{Eq:in_current_dens}
    2n^+\beta(x) = \frac{j_0}{e}.
\end{equation}
Then, following the second symmetric function, (Eqs. \ref{Eq:small_f_photon} and \ref{Eq:boundary_photon}) 
\begin{equation}\label{Eq:photon_sym_bound}
    f^+_i = f^-_i \ (i = 1, 2, ..., m),
\end{equation}
for each value of $\epsilon_\gamma$.\\

\indent The boundary at $x = H$ is formulated to be an free boundary that ensures $E_{||}$ smoothly vanishes as it approaches $H$. To accomplish this, any inwardly propagating particle - in this scenario the $e^-$'s - should not enter the gap from $x > H$. Combining this condition with (Eq. \ref{Eq:current_density}) at $x = H$ results with
\begin{equation}\label{Eq:out_part_H}
    n^-(H) = 0 -> n^+\beta(H) = \frac{j_0}{e}.
\end{equation}
Additionally, the charge density distribution must remain continuous at the outer boundary through (Eqs. \ref{Eq:Poisson_reduc_GJ} and \ref{Eq:out_part_H}) to recover
\begin{equation}\label{Eq:charge_dens_dist}
    \frac{1}{4\pi}\frac{dE_{||}}{dx} = j_0\beta(x) - Ax.
\end{equation}
Similar to the inner boundary, no $\gamma$-ray photons may come into the gap from the outside (all up-scattered photons must be created inside the gap), represented mathematically via
\begin{equation}\label{Eq:photons_create_gap}
    f_i^- = 0 \ (i = 1,2,...,m) \ at \ x=H.
\end{equation}
\\

\section{Simulating $e^\pm$ Cascade}
\indent From the above derivations, there is now $(2m + 5)$ total boundary conditions for $(2m + 3)$ total unknown functions for $E_{||}(x)$, $\Gamma(x)$ (which is contained in the $\beta(x)$ function), $n^+$, and $f_i^\pm(x)$ for $(i = 1,2,...,m)$ and two uncalculated constants $H$ and $j_0$. These are formed from the conditions $E(H) = 0$ and $E_x^{'}(H) = 0$, so $j_0$ now plays the role as an eigenvalue of the boundary value problem.\\

\indent It is sufficient for the investigation into the pair production to consider $\gamma$-ray photons satisfying 
\begin{equation}\label{Eq:photon_energy}
    \epsilon_\gamma > \frac{2}{\epsilon_s(1 - \mu)},
\end{equation}
which is a dimensionless energy parameter. Below the minimum energy threshold of $\epsilon_{min} = \epsilon_{max}^{-1}$, photons fail to contribute to pair production. The lowest energy bin may be defined as $\beta_0m_ec^2$ with $\beta_0 = \epsilon_{max}^{-1}$. These may be chosen based on the study at hand and will be numerically defined in the following section. The underlying goal is then to seek solutions to the boundary conditions by way of the shooting method of solving boundary value problems. This method involves taking the boundary value problem and reducing it to an initial value problem at different conditions until the solution that satisfies the original problem is met. For this study, the shooting algorithm begins at $x = 0$ and finishes at the outer boundary $x = H$.\\

\indent For a realistic model of a BH, the current flowing along each field line is determined more by the global system physics rather than the micro-scale physics inside the gap. \citep{24} shows that the load connecting wind/jet to the unipolar induction is the reasoning behind this. We define the ``null surface'' to be where the gap center is located and where $\rho_{GJ}$ vanishes. This definition imposes a symmetrical center point in the gap, and allows $j_0$ to be treated as an eigenvalue in a standard boundary value problem (facilitating the shooting method) and focusing on finding $j_0$ with the outer bound of $x = +H$. Following this, the shooting method scheme can be listed as follows:
\begin{itemize}
    \item Very small values of $j_0$ (with $|n^+ - n^-|$ is also very small), (Eq. \ref{Eq:Poisson_reduc_GJ}) can be approximated with
    \begin{equation}\label{Eq:shoot_E_par}
        \frac{dE_{||}}{dx} \approx E_{||}(0) - 4\pi Ax^2.
    \end{equation}
    \item The above quadratic form function cannot solve 
    \begin{equation}\label{Eq:shoot_quad}
        \frac{1}{4\pi}\frac{dE_{||}}{dx} = j_0\beta(x = H) - Ax|_H = 0.
    \end{equation}
    \item As $j_0$ increases, the RHS will also increase (monotonically) with $x$ in (Eq. \ref{Eq:Poisson_reduc_GJ})
    \begin{equation}\label{Eq:shoot_poisson}
        \frac{dE_{||}}{dx} = 4\pi[e(n^+ - n^-) - Ax].
    \end{equation}
    \item As $j_0$ increases more, $\frac{dE_{||}}{d(x=H)}$ continues to decrease until it vanishes at some $j_{cr}(U_b,\alpha)$. 
    \item Above $\j_{cr}$, the second condition can not be satisfied, no matter what initial conditions are given.
\end{itemize}
To summarize thus far: the system seeks a solution that satisfies the second condition by adjusting $j_0$ to $j_{cr}(U_b, \alpha)$. This resulting system is then molded by $29$ different equations for $31$ different unknown variables which are then integrated under $31$ boundary conditions. A table summarizing these boundary conditions can be seen in Appendix C.\\

\indent To study the $e^\pm$ cascade seen in the magnetospheres of KBHs, a $C^{++}$ code \citep{7,20} is used to solve the system of equations for a given set of initial conditions. To conduct this analysis, a set of initial conditions was set and solved for. This control initial condition set can be found in Appendix B8. From here, a single parameter is incrementally changed (``swept’’) and the new system is then solved. This generates a new set of initial conditions the code then solves for and the process repeats; set, change, solve, set. This incremental change must be small enough that the code can start its process using the previous set of initial conditions and the new parameter.\\

\indent This code relies on the shooting method of solving a boundary value problem (BVP) using a set of initial parameters defined at the top of the code (gap width, Lorentz factors, energy bins, etc.). The shooting method takes the multi-dimensional BVP and simplifies it to a single initial value problem (IVP). From the IVP, the code takes the solution to the equations at one boundary ($x = 0$) and attempts to find the solution that solves the equations at the other bound ($x = H$) by ``shooting’’ different solution guesses to the boundary. Once the solution is found, namely $j_0$, the remaining parameters are solved for and the parameter under analysis is incrementally changed, and the code solves the system again. If the change in the swept parameter is too big, the guess of solution will be too far off to solve, and the system of equations will collapse. \\

\indent While this code accurately solves the mathematics within the gap, the scope of the code stops there. Despite the importance of the solutions found in this gap, it is also important to visualize how this plasma fuels the jets at the poles. Various studies, including ones done by \citep{31,32,33,34} use PIC codes to model the system globally, from soft photon bath to jets. These studies are just as important as they allow us to study the state of plasma and gap and view their effects on the jets themselves; which we view through observational astronomy.

\section{Analysis of Changing Flux Model}
\indent One of the main goals of this study is to alter the flux model used from the simplified model seen in previous studies \citep{20} to a more realistic model motivated by simulations \citep{31}. The old flux expression, $sin(\theta)^2$, represents a toy model of accretion. Caveat is “pathology” at large distances, represented by a diagonal line at large $r$. The updated model, $(1- cos(\theta))$, has no caveat present. Physically, this model represents sites with no accretion (MW type systems), accretion from a thick, hot disk (AGN systems), and far away, thin disks. Pictorially, we can see the null surface reaching outward from the surface of the KBH for the toy model. Mathematically, the difference between the flux models $\theta$ dependence of $A_{\theta} \equiv \partial_r \rho_{GJ}$. This difference stems from the expansions coefficient's ($A$) distribution of field around the null surface, $x = 0$ (within first order). For the updated model, the null surfaces stretch at equal latitudes around the BH. This new flux model more accurately represents the magnetic field structure – split monopole. Simulations show the split monopole configuration forms near KBHs, so it is a natural force free configuration. A table of the initial conditions (ICs) used in this paper are found in Appendix B.\\

\begin{figure*}[h]
    \centering
    \begin{subfigure}
        \centering
        \includegraphics[scale=0.3]{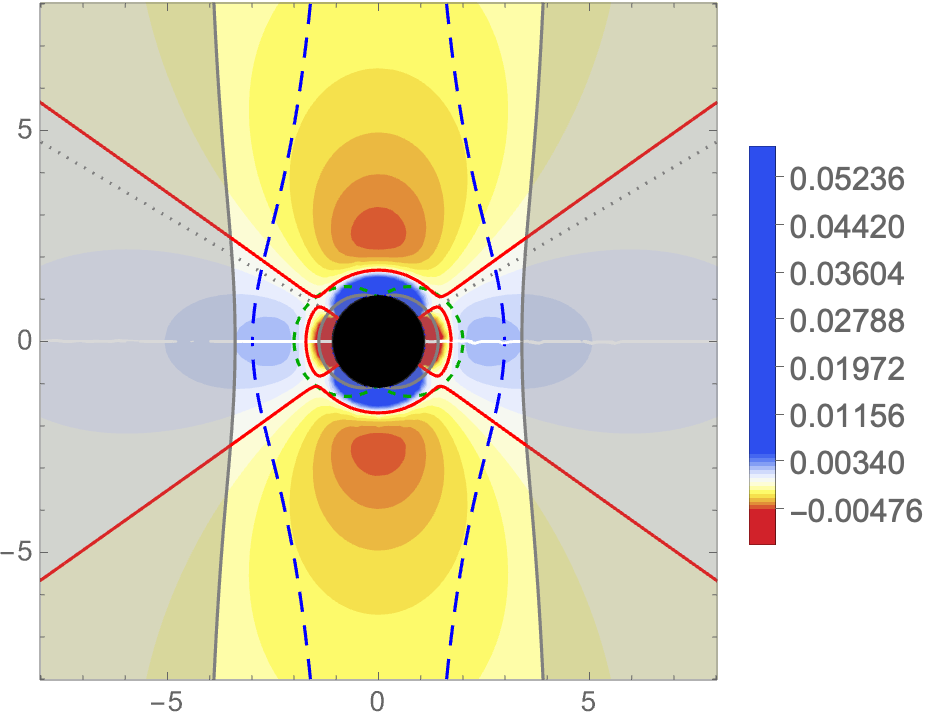}
    \end{subfigure}
    \begin{subfigure}
        \centering 
        \includegraphics[scale=0.3]{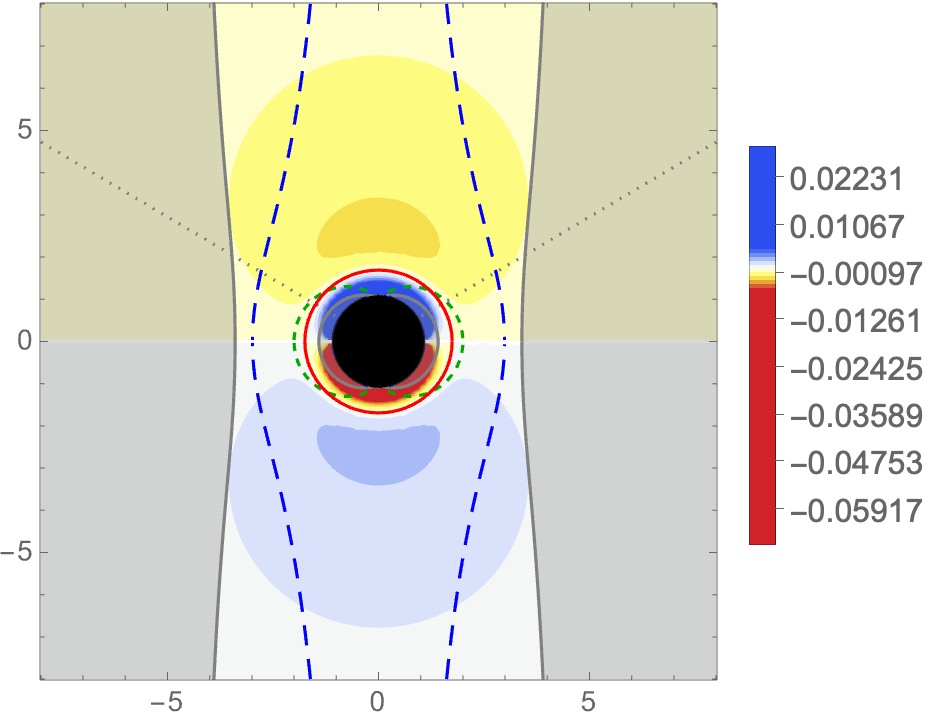}
    \end{subfigure}
    \vskip\baselineskip
    \begin{subfigure}
        \centering
        \includegraphics[scale=0.3]{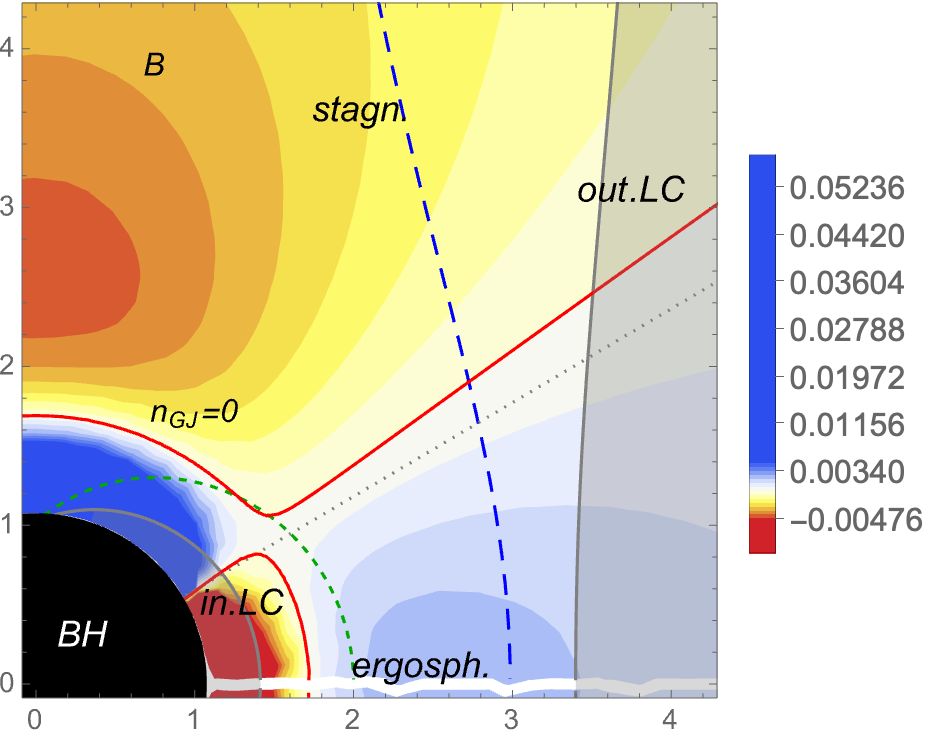}
    \end{subfigure}
    \begin{subfigure}
        \centering 
        \includegraphics[scale=0.3]{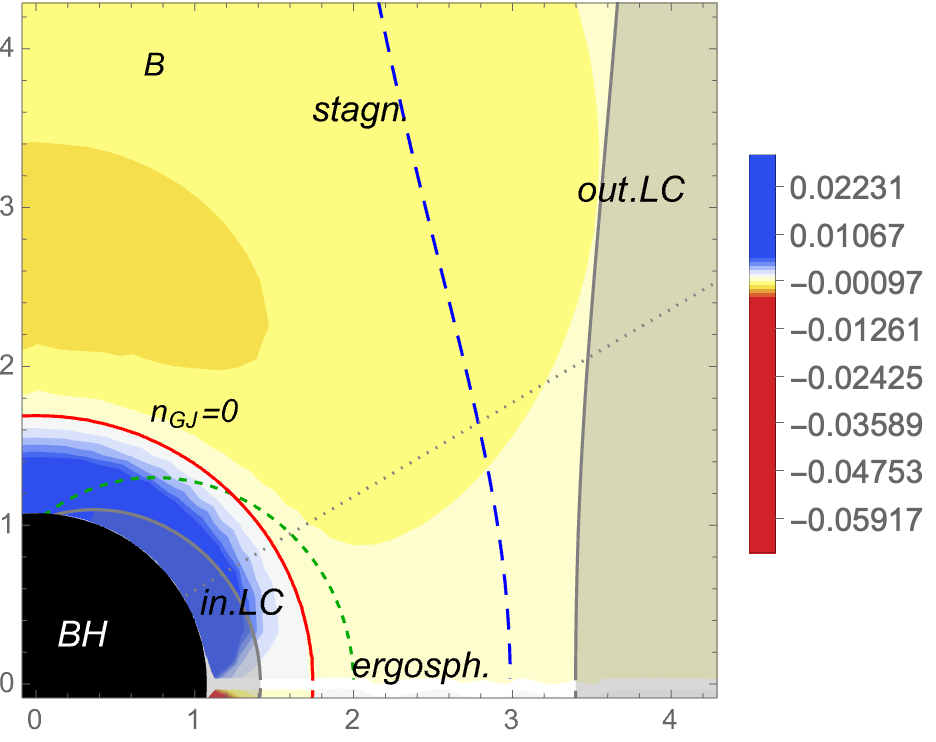}
    \end{subfigure}
    \caption{Visual comparison of the charge density environment of the KBH. $sin(\theta)^2$ is shown on the left and $(1 - cos(\theta))$ on the right. In arbitrary units, $\rho_{GJ}$ is plotted around the KBH with blue signifying positive density and red negative. The new model shows a less layered distribution of charge with lesser magnitude areas being seen outside the null surface, away from the KBH. The differences in the null surface distribution are seen in the red solid line to follow the less complex distribution of charge as the new model shows a single shell around the KBH while the old model (left) shows a farther reaching trend.}
    \label{Fig:Flux_Comp_RGJ_Solutions}
\end{figure*}

\begin{figure*}[h]
    \centering
    \begin{subfigure}
        \centering
        \includegraphics[scale=0.3]{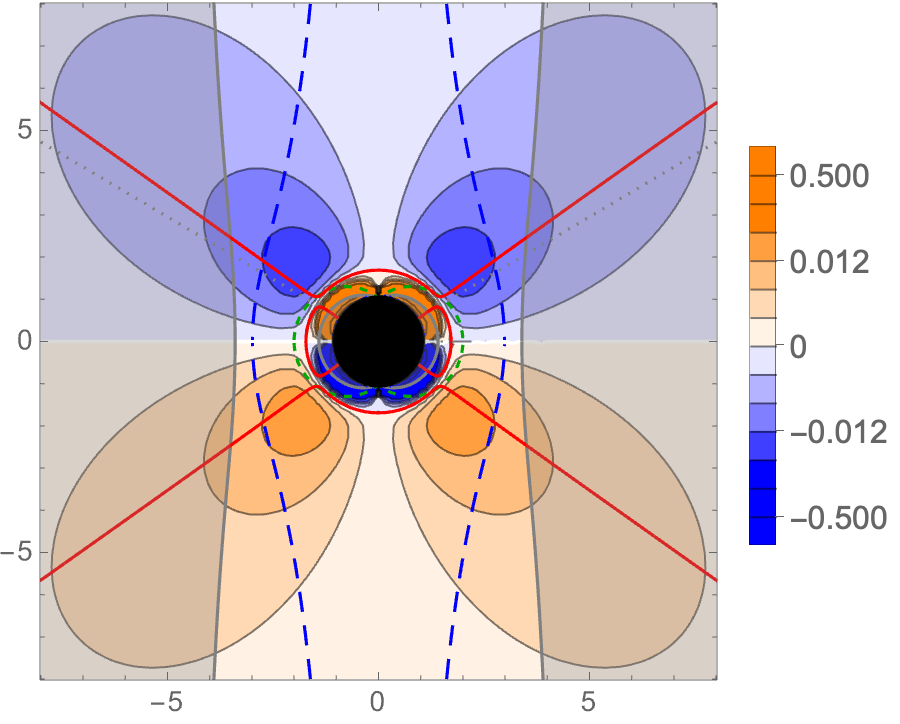}
    \end{subfigure}
    \begin{subfigure}
        \centering 
        \includegraphics[scale=0.3]{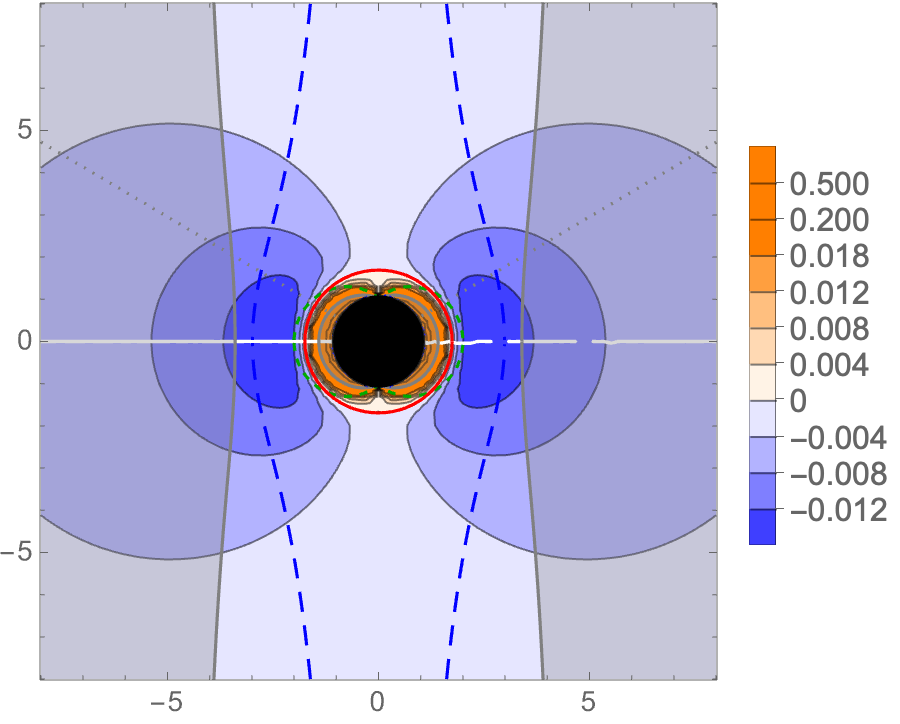}
    \end{subfigure}
    \vskip\baselineskip
    \begin{subfigure}
        \centering
        \includegraphics[scale=0.3]{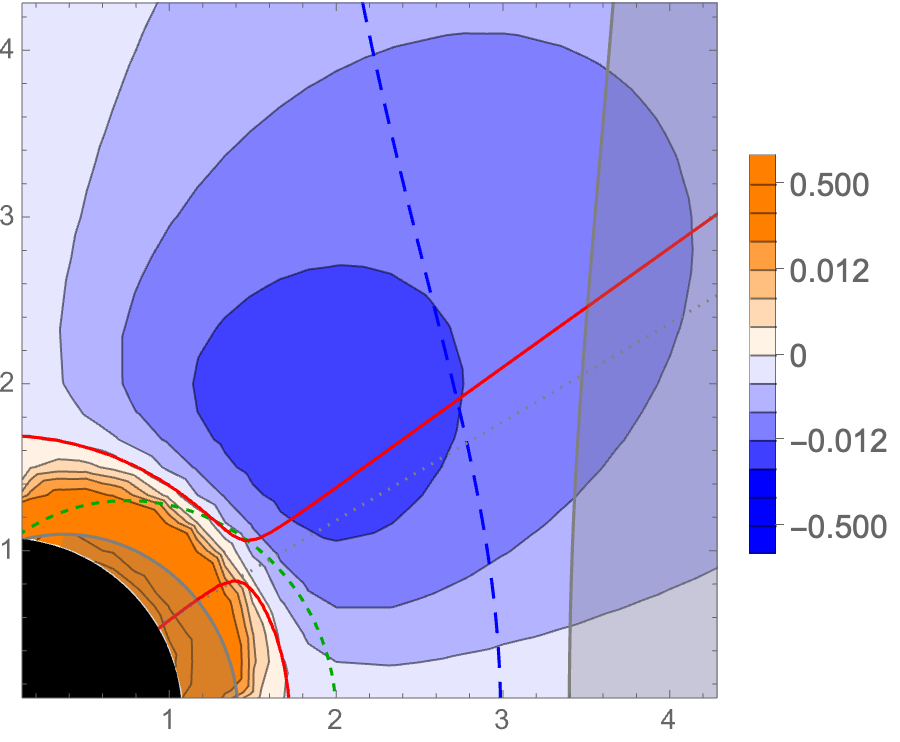}
    \end{subfigure}
    \begin{subfigure}
        \centering 
        \includegraphics[scale=0.3]{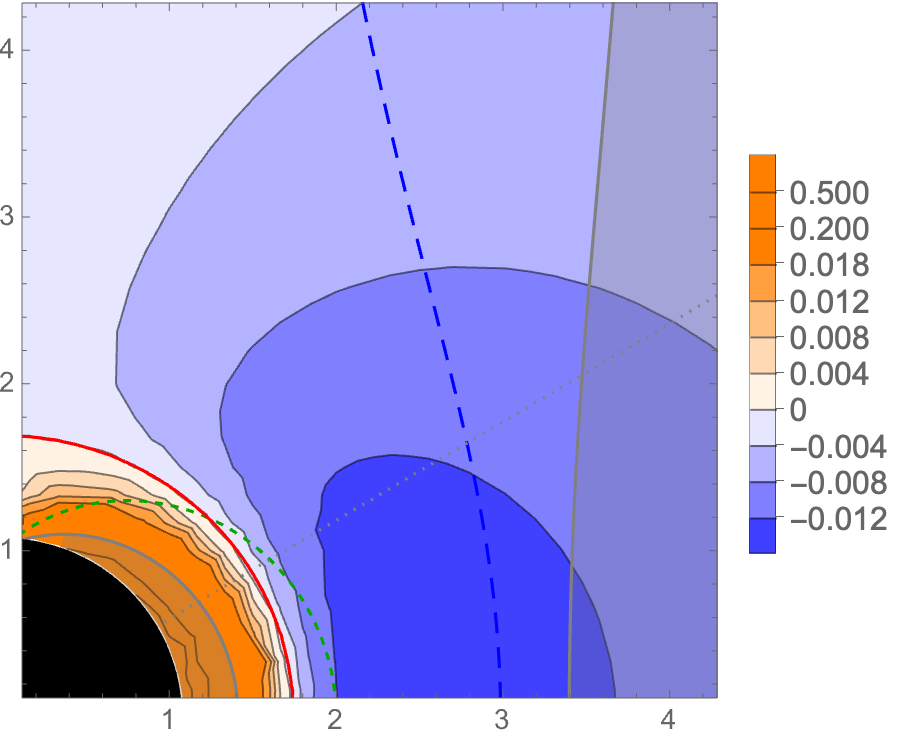}
    \end{subfigure}
    \caption{Visual comparison of the parallel electric field of the KBH. $sin(\theta)^2$ is shown on the left and $(1 - cos(\theta))$ on the right. In arbitrary units, $\vec{E}_{||}$ is plotted around the KBH with green signifying positive field and red negative. Again, the new model shows a less complex distribution of field. The same contours are plotted for each model, and with this we see the relative bunching of the field in the old model as well as the less circular shapes of the new model's contours as it folds around the event horizon.}
    \label{Fig:Flux_Comp_Ep_Solutions}
\end{figure*}

\indent The differences in the flux model can be examined through global visualizations of the solutions to many of the equations presented in the section above. Looking at distinct theoretical solutions for $\rho_{GJ}$, stagnation point (flow of in flowing or out flowing material is zero), light cylinders (regions of warped spacetime so severs that there is no sub-luminal path out), and ergosphere (region in which a observer must rotate with the KBH and can no longer be a stationary observer), initial differences in the models can be seen, but are not complete. Building on these solutions, we can view more differences in the flux models by the plasma density $\rho_{GJ}$ (Fig. \ref{Fig:Flux_Comp_RGJ_Solutions}) and the parallel electric field (Fig. \ref{Fig:Flux_Comp_Ep_Solutions}). These plots more greatly illustrate the major differences in field and plasma density distribution. We see two very distinct model structures in both quantities; in structure, magnitude, and distribution. The effects these differences have on the structure of the gap will be examined in the following subsections.

\subsection{Flux Transition}
\indent To study this model, the code must first be set up for the new flux conditions. As stated above, the code must sweep a parameter from the control initial conditions to the desired point, as opposed to plugging in the new model directly. To achieve this, the flux models are written in a summation form with each other as a function of percentage, with $\eta$ representing the percent of the old model and $\iota$ representing the percent of the new model: 
\begin{equation}\label{Eq:Flux_Superposition}
    \Psi(\theta) = \eta(sin(\theta)^2) + \iota(1 – cos(\theta)).
\end{equation}
where we define $\iota = \eta – 1$ to restrict the free variables to only one. The code begins with $\eta = 1$ and incrementally decreases until $\eta = 0$, where the new flux model is fully implemented. This change of flux is done at the axis of rotation of the KBH $\theta = 0$. \\

\indent Plotting the three main gap characteristics viewed in this study (half gap width, Lorentz factor at gap center, and outgoing particle flux), we can examine the transition of fluxes at the polar cap of the KBH. As a note: the plots shown in this paper that are labeled ``normalized’’ are normalized to the initial value in that specific parameter sweep. This allows a progression trend to be seen with respect to the starting point.\\

\begin{figure}
    \centering
    \begin{subfigure}
        \centering
        \includegraphics[scale=0.5]{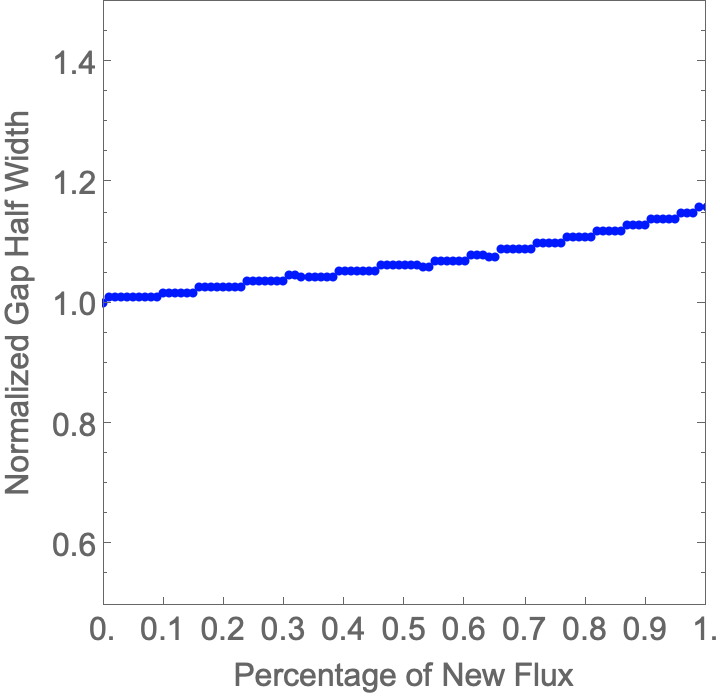}
    \end{subfigure}
    \begin{subfigure}
        \centering 
        \includegraphics[scale=0.5]{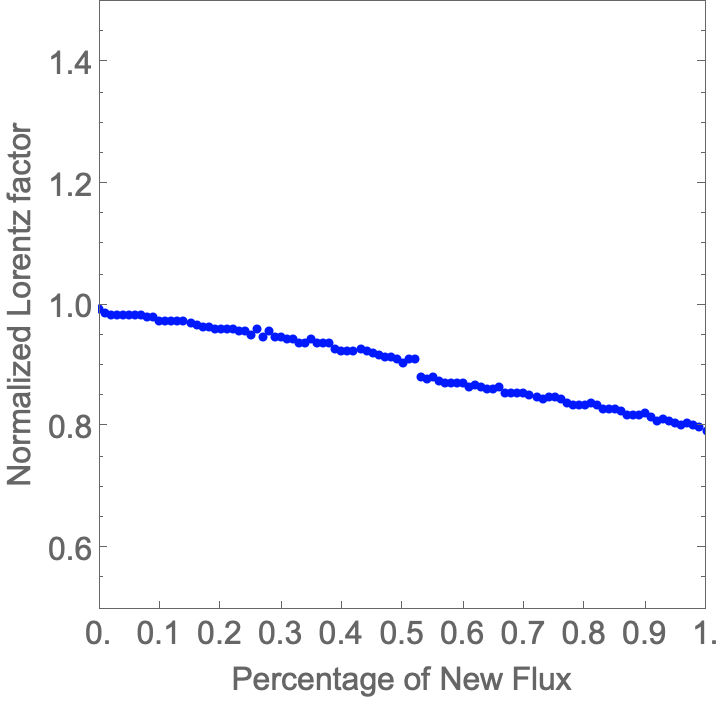}
    \end{subfigure}
    \begin{subfigure}
        \centering 
        \includegraphics[scale=0.5]{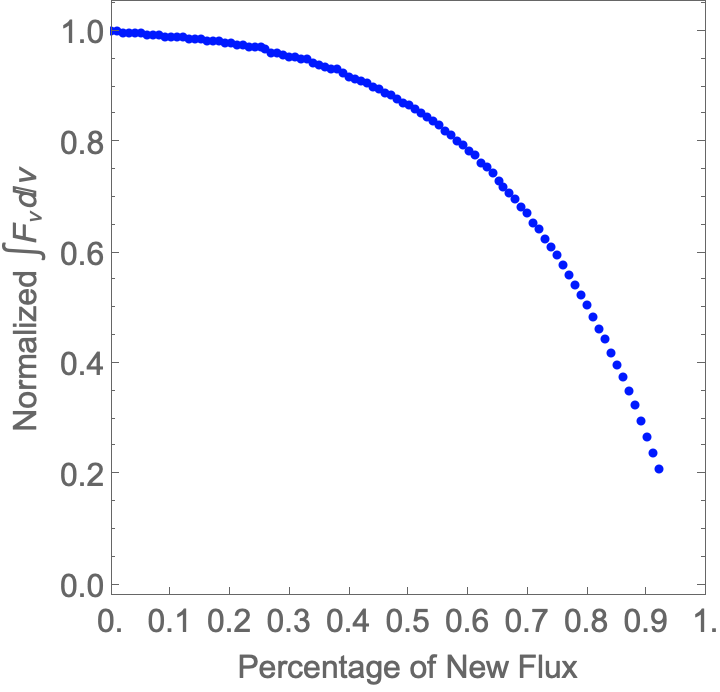}
    \end{subfigure}
    \caption{The transition between the two flux models can be viewed as a function of percentage of the old model. Though these figures only show the axis of rotation, it does give a insight to how the gap will change over all angles. We see the gap widen as the new model takes over, this widening leads to more space for particles to decelerate. This shows up in the Lorentz factor of the particles steadily decreasing. With this decrease in velocity, the outgoing flux follows, as seen in the exponential decrease of outgoing flux.} 
    \label{Fig:Norm_Flux_Trans}
\end{figure}

\indent We can immediately see in (Fig. \ref{Fig:Norm_Flux_Trans}) the size of the gap gradually increases as the flux shifts to the new, realistic model. With the increase in gap size, there is also a decrease of Lorentz factor within the center of the gap. This can be from the expanded space in which particles decrease motion and loose energy. The most drastic change in trend is shown in the normalized outgoing particle flux, where we see a trend that leads to almost $100\%$ decrease in flux at the polar cap. The loss of energy in particles from the Lorentz trend hints that they can no longer scatter of photons, decreasing the flux.  While these trends can show general behavior or trend, they are not indicative of the while system, as they are only at $\theta = 0$.\\

\begin{figure*}
    \centering
    \begin{subfigure}
        \centering
        \includegraphics[scale=0.45]{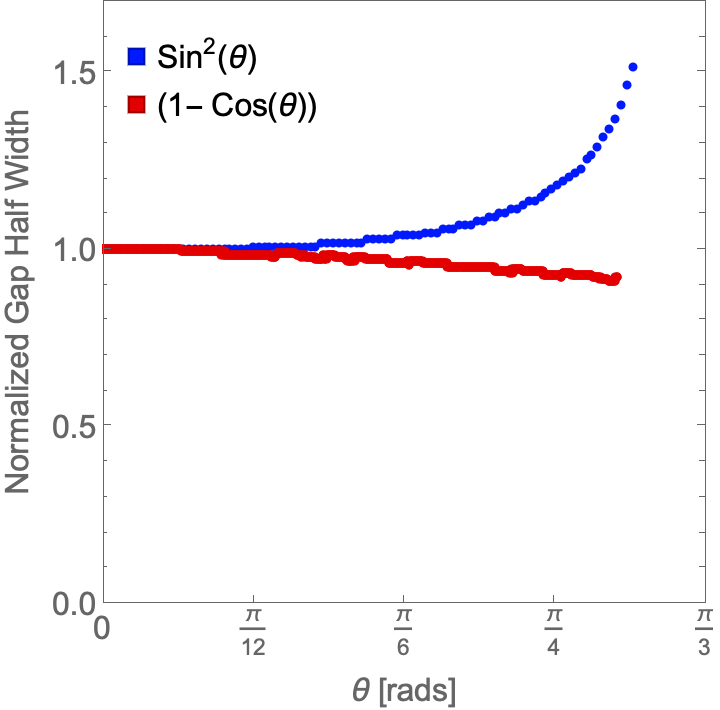}
    \end{subfigure}
    \begin{subfigure}
        \centering
        \includegraphics[scale=0.45]{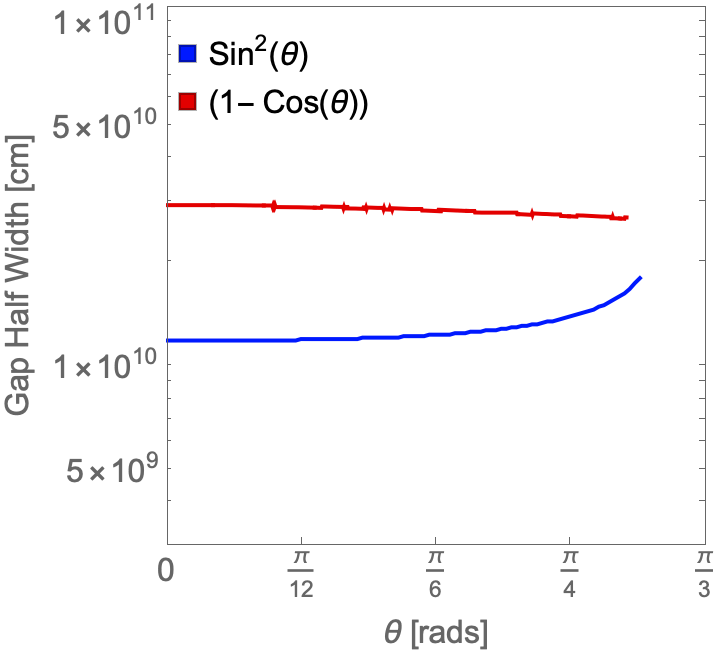}
    \end{subfigure}
    \begin{subfigure}
        \centering 
        \includegraphics[scale=0.45]{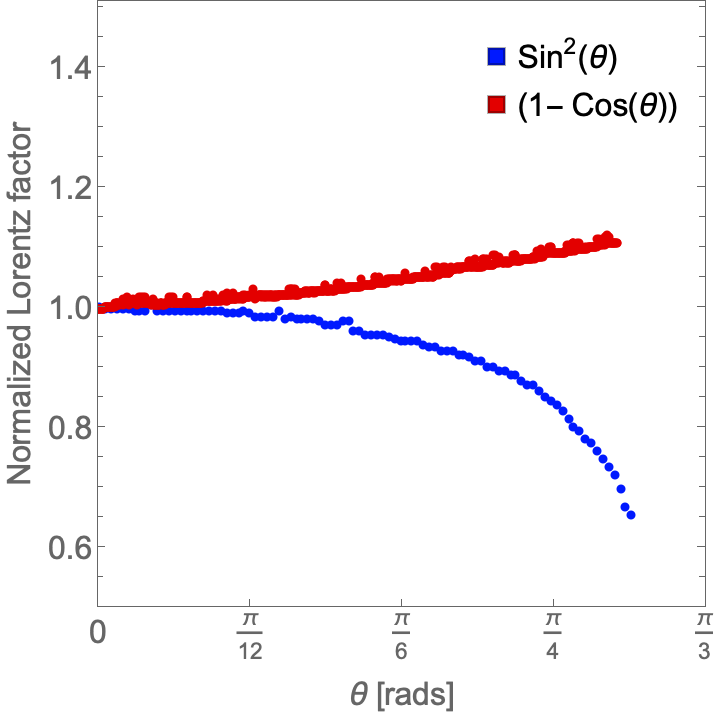}
    \end{subfigure}
    \begin{subfigure}
        \centering 
        \includegraphics[scale=0.45]{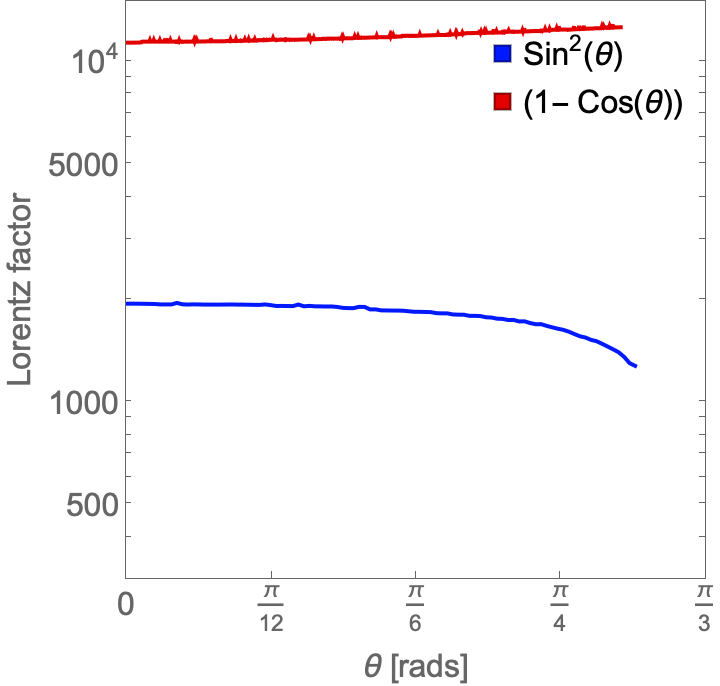}
    \end{subfigure}
    \begin{subfigure}
        \centering 
        \includegraphics[scale=0.45]{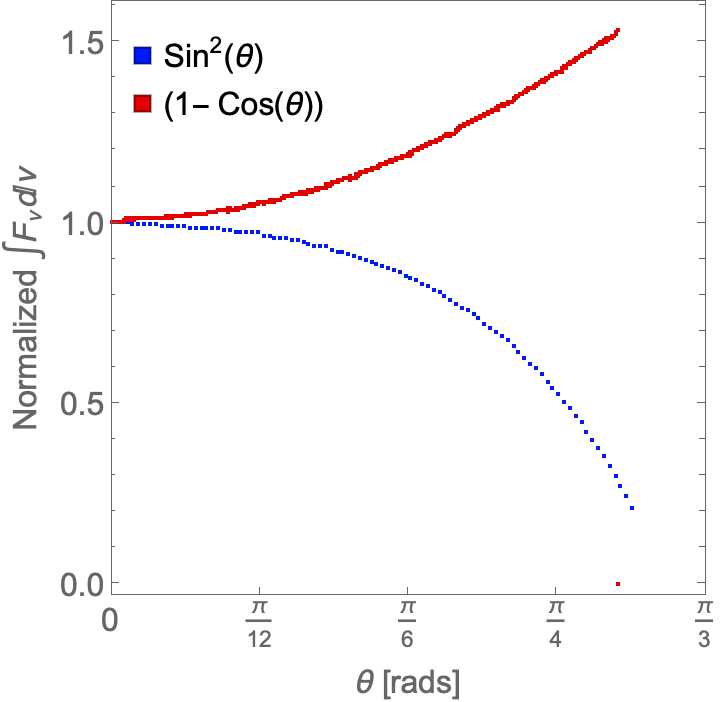}
    \end{subfigure}
    \begin{subfigure}
        \centering 
        \includegraphics[scale=0.45]{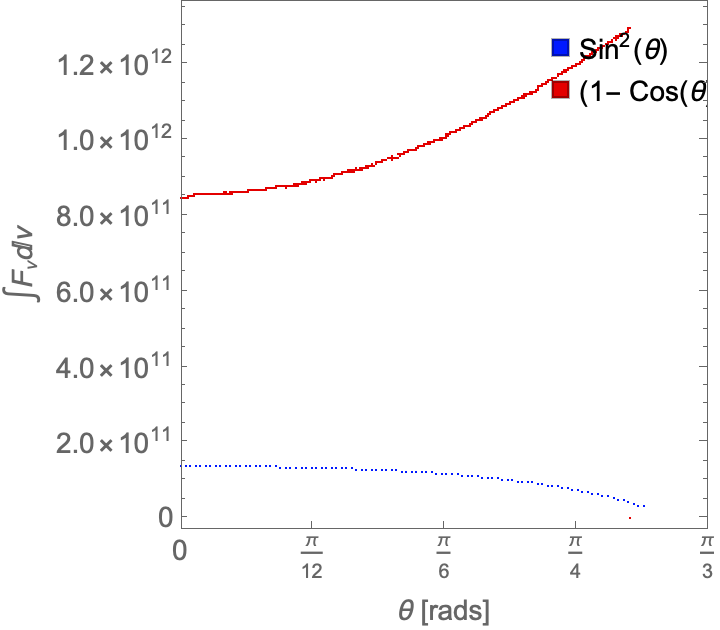}
    \end{subfigure}
    \begin{subfigure}
        \centering 
        \includegraphics[scale=0.45]{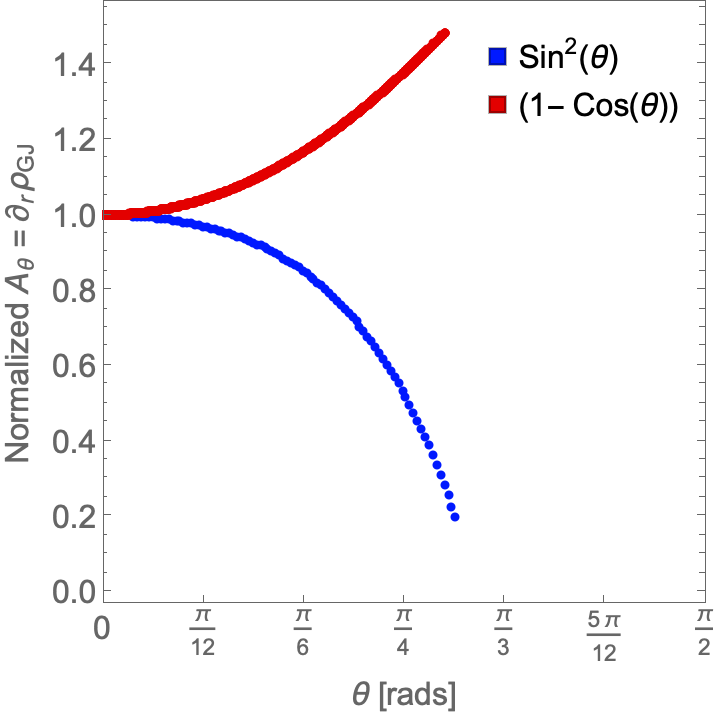}
    \end{subfigure}
    \begin{subfigure}
        \centering 
        \includegraphics[scale=0.4]{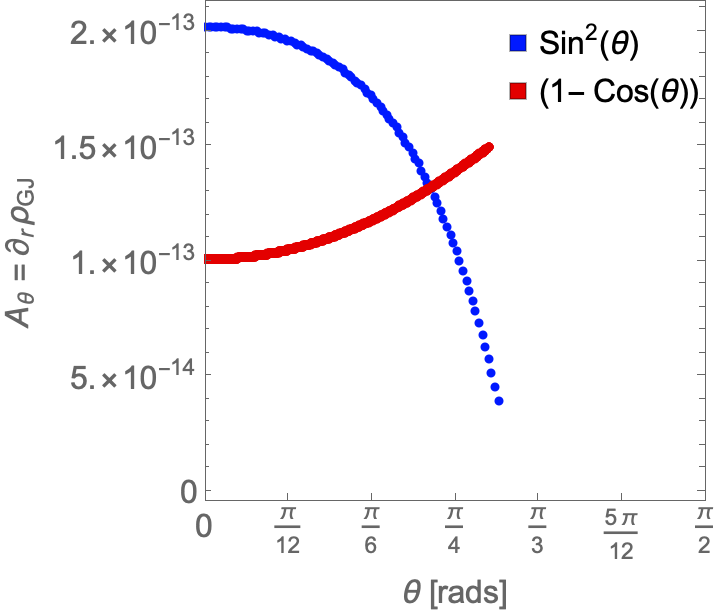}
    \end{subfigure}
    \begin{subfigure}
        \centering 
        \includegraphics[scale=0.45]{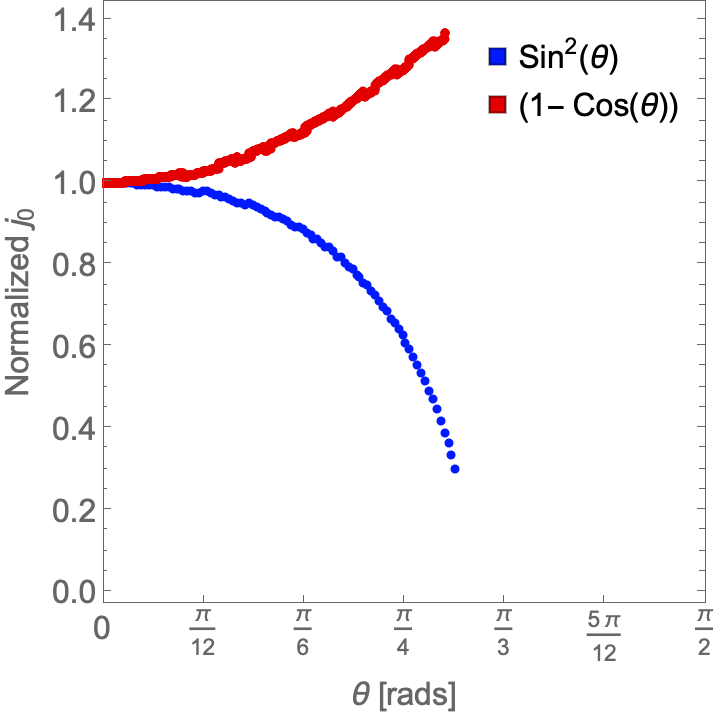}
    \end{subfigure}
    \begin{subfigure}
        \centering 
        \includegraphics[scale=0.45]{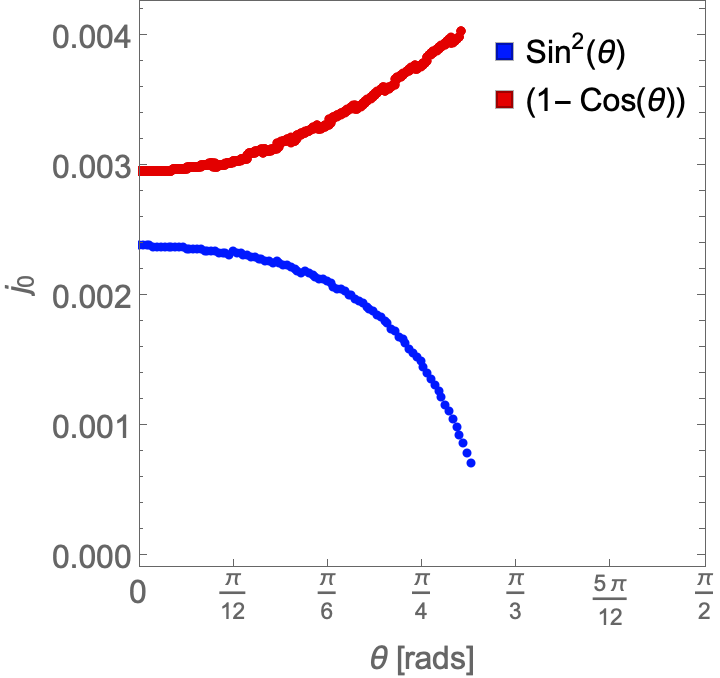}
    \end{subfigure}
    \caption{After the new flux model is fully integrated, the two expressions can be directly compared at the axis of rotation ($\theta = 0$). The normalized trends are shown on the right while the unnormalized data is plotted on the left. As expected, the direction of the new model's trend is opposite of that of the old model. The new model also shows less severity as the model sweeps from the pole to the equator.}
    \label{Fig:Flux_Comp_5_Values}
\end{figure*}

\indent Viewing the comparison of the old model to the new model (Fig. \ref{Fig:Flux_Comp_5_Values}) in the same KBH system (that is only the flux has changed from the control ICs) gives a more complete view of what should be expected with the split monopole model. Immediately, it can be seen that all three plots show an oppositely directed trend, which is what is expected in this switch. It can also be noted the lack of severity in the trends when compared to the $sin(\theta)^2$ model. This depicts a more evenly structured gap and a less drastic change in emission based on viewing angle to the axis of rotation. \\

\indent The observed opposite trends for the quasi-spherical gas distribution studied here and the thin accretion disk studied in a previous paper \citep{20} indicates the importance of the magnetic field geometry. The former (quasi-spherical) case show the highest cascade efficiency at high latitudes, toward the jet axis. The latter, in contrast, indicates highest efficiency around the jet-disk interface. This can be a direct prediction of our study and a possible observational diagnostic of the field geometry at astrophysical sources.

\clearpage
\newpage

\subsection{Varying Mass}
\begin{figure*}
    \centering
    \begin{subfigure}
        \centering
        \includegraphics[scale=0.6]{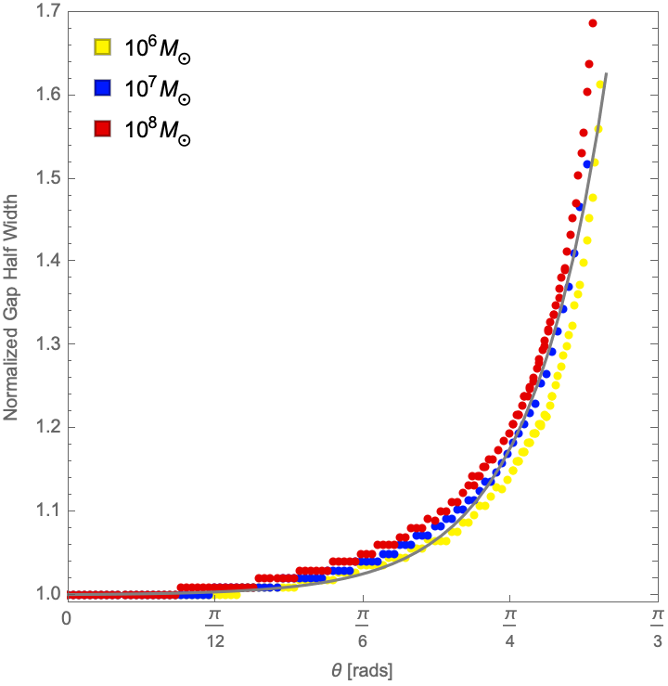}
    \end{subfigure}
    \begin{subfigure}
        \centering 
        \includegraphics[scale=0.6]{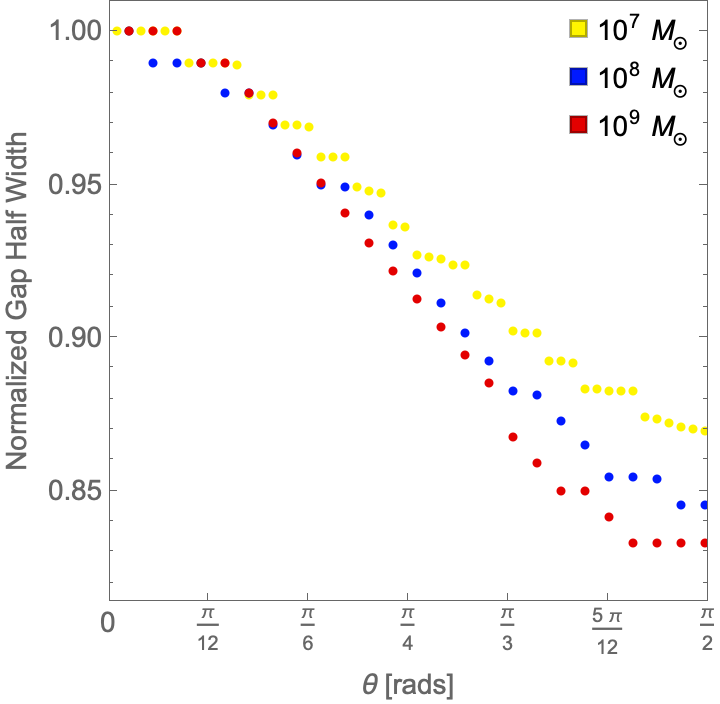}
    \end{subfigure}
    \vskip\baselineskip
    \begin{subfigure}
        \centering 
        \includegraphics[scale=0.6]{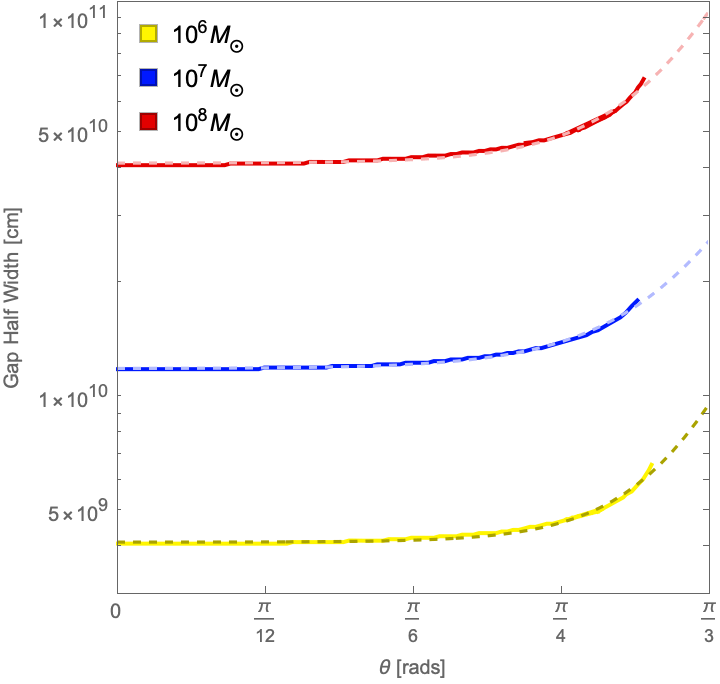}
    \end{subfigure}
    \begin{subfigure}
        \centering 
        \includegraphics[scale=0.6]{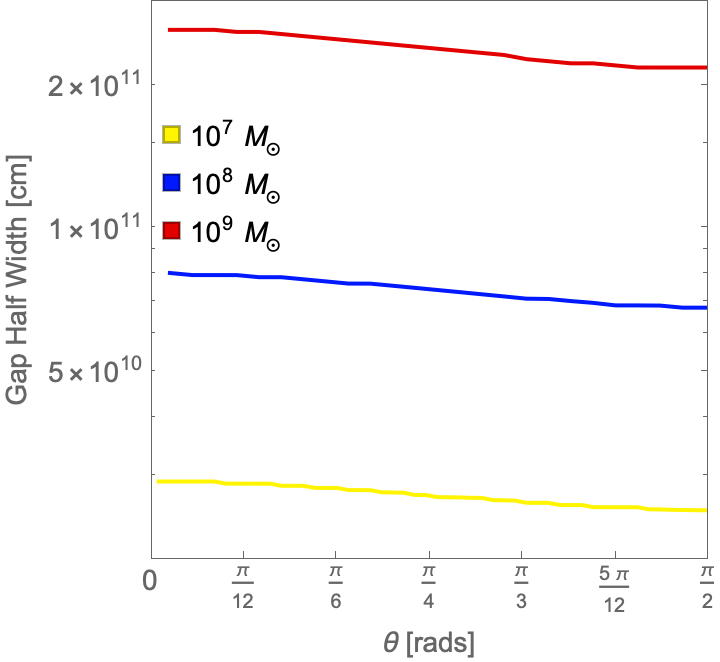}
    \end{subfigure}
    \caption{Comparison of gap widths in old and new models (left and right panels, respectively). The old model ($sin^2(\theta))$ is plotted on the left with the normalized trend above the unnormalized trends. We can immediately see the opposite and more complex nature of the $(1- cos(\theta))$ model. We also note that the old models trends are more bunched in the normalized plots while the new model has variance.} 
    \label{Fig:Gap_Wdiths_Flux_Comp}
\end{figure*}

\indent To fully investigate the differences in the flux models, direct comparisons from parameters studied in the old flux model must be made. For this study, KBH mass was varied and then examined as a function of angle $\theta$. For the old flux model, data from \citep{20} was used to study KBH's of $10^6 M_{\odot}$, $10^ 7 M_{\odot}$, and $10^8 M_{\odot}$. For the updated mass model,  $10^7 M_{\odot}$, $10^8 M_{\odot}$, and $10^9 M_{\odot}$ was used. To begin, the half gap width of the gap is examined in (Fig. \ref{Fig:Gap_Wdiths_Flux_Comp}). In the unnormalized plots, the higher the mass, the larger the gap is in a general sense. Continuing in the unnormalized trends, not only is there no drastic slope change as the gap approaches the equator, but the direction of the trend itself opposite, with the new model constantly decreasing. Viewing the trend as a ratio from the pole to the equator, we see many differences in the models. Starting with the obvious, instead of a increasing exponential we see a decreasing quasi-linear relation, with plateaus at the x-axis limits. We also see more definite spacing between the masses as opposed to the tight grouping of the $sin(\theta)$ model. \\

\begin{figure*}
    \centering
    \begin{subfigure}
        \centering
        \includegraphics[scale=0.6]{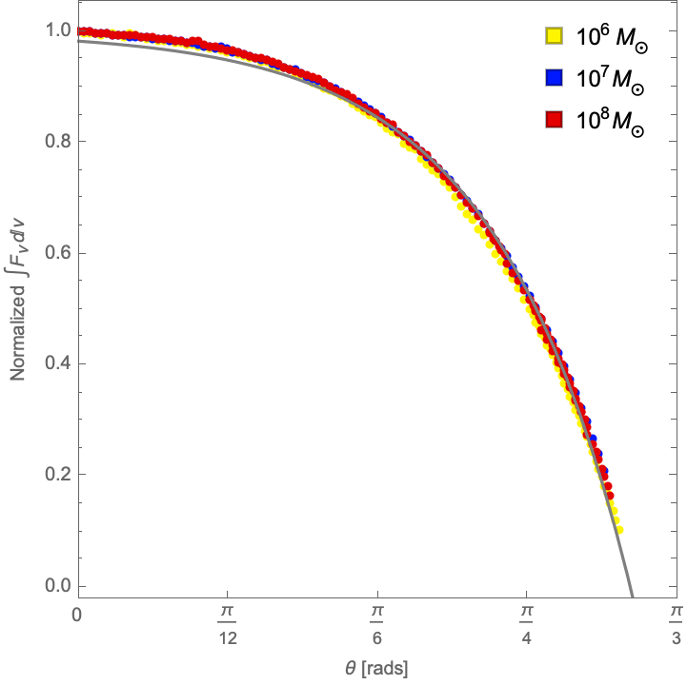}
    \end{subfigure}
    \begin{subfigure}
        \centering 
        \includegraphics[scale=0.6]{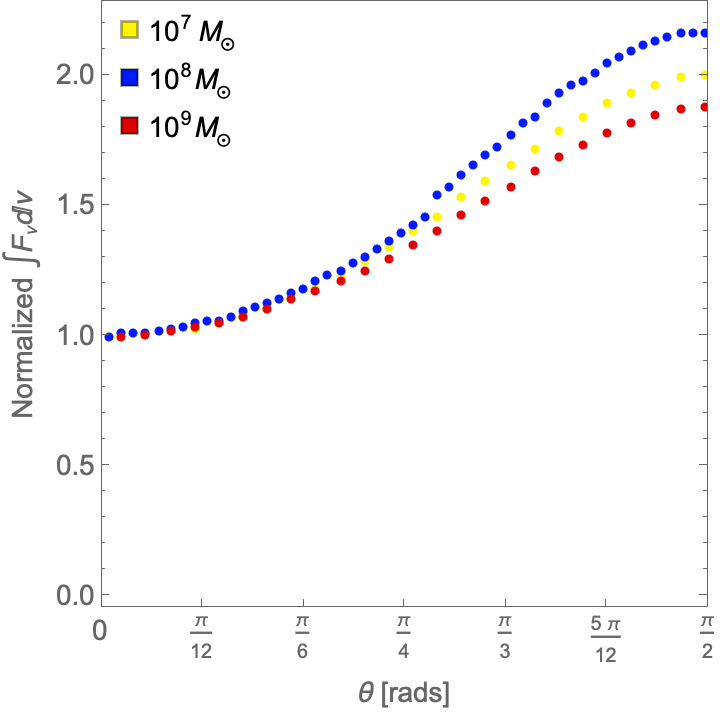}
    \end{subfigure}
    \vskip\baselineskip
    \begin{subfigure}
        \centering 
        \includegraphics[scale=0.6]{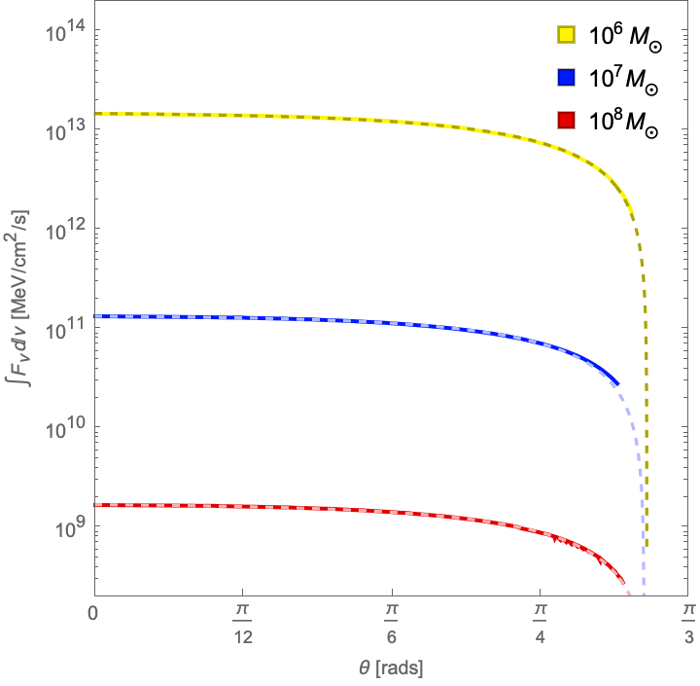}
    \end{subfigure}
    \begin{subfigure}
        \centering 
        \includegraphics[scale=0.6]{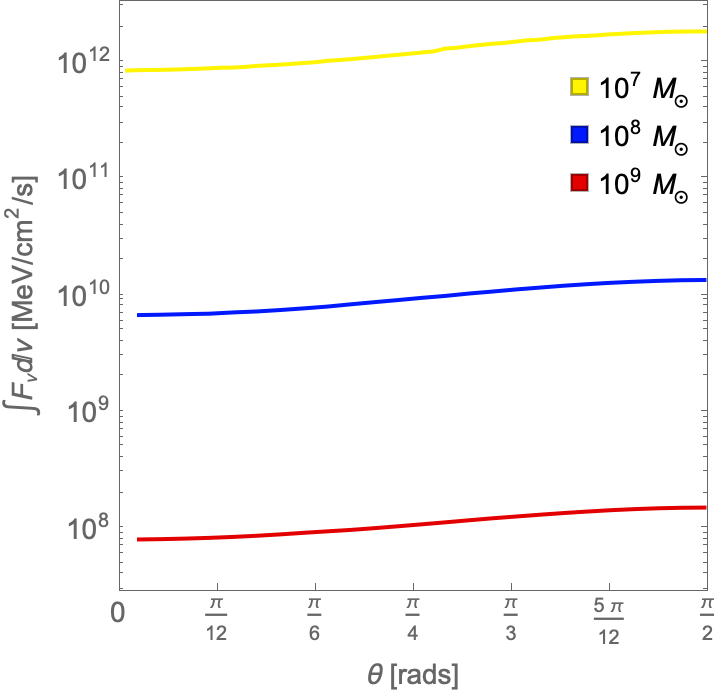}
    \end{subfigure}
    \caption{Comparison of $\gamma$-ray fluxes in old and new models (left and right panels, respectively). The old model ($sin^2(\theta))$ is plotted on the left with the normalized trend above the unnormalized trends. We see the same behaviors and inter-model relationships as before. Interestingly, the total $\gamma$-ray flux increases toward the equator, in contrast to the old model.} 
    \label{Fig:Out_Flux_Flux_Comp}
\end{figure*}

\indent The outgoing flux is discussed (Fig. \ref{Fig:Out_Flux_Flux_Comp}) next. We see for both models that the higher the mass of the KBH, the less outgoing flux overall. The updated model sees as small general increase within the mass trends versus the exponential decrease of the toy model, in the unnormalized figures. Under normalization, the old model again displays a bunched exponential tend with the decrease of flux as the angle increases. The new model displays a bunch to spread relationships between the masses, with the lowest mass being between the highest and lowest mass. The trend shows a $sin(x^2)$ function, see Appendix D.\\

\begin{figure*}
    \centering
    \begin{subfigure}
        \centering
        \includegraphics[scale=0.6]{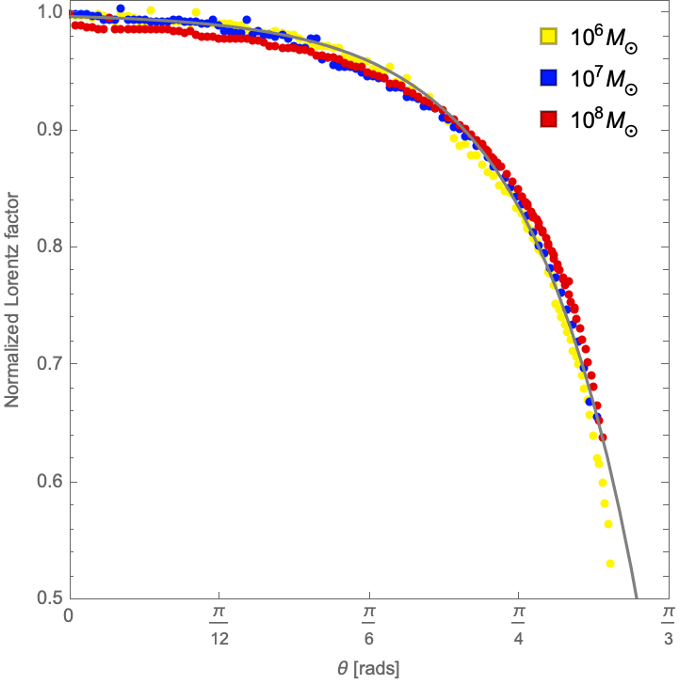}
    \end{subfigure}
    \begin{subfigure}
        \centering 
        \includegraphics[scale=0.6]{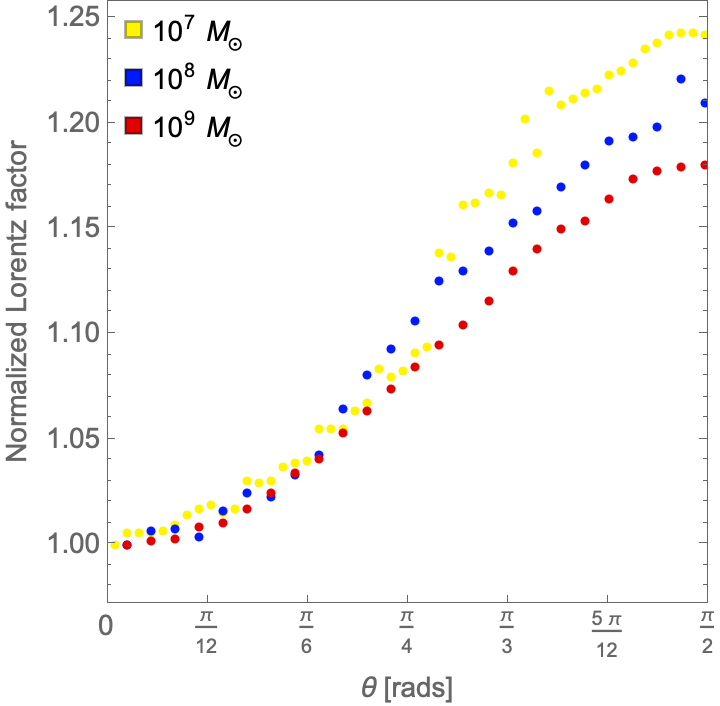}
    \end{subfigure}
    \vskip\baselineskip
    \begin{subfigure}
        \centering 
        \includegraphics[scale=0.6]{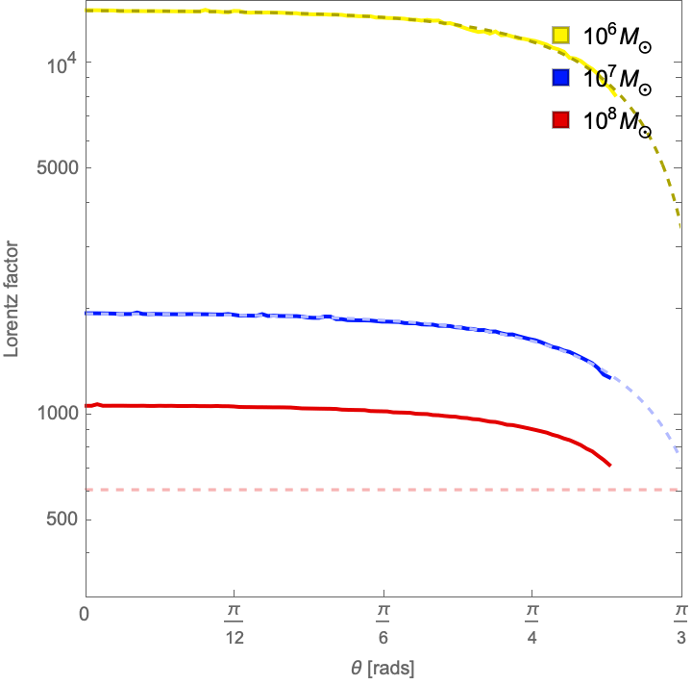}
    \end{subfigure}
    \begin{subfigure}
        \centering 
        \includegraphics[scale=0.6]{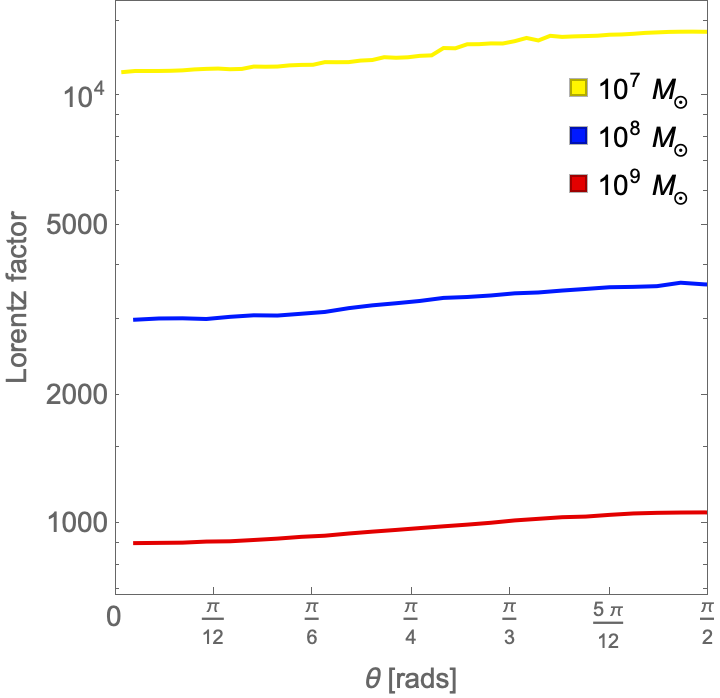}
    \end{subfigure}
    \caption{Comparison of peak Lorentz factor in old and new models (left and right panels, respectively). The old model ($sin^2(\theta))$ is plotted on the left with the normalized trend above the unnormalized trends. We again see the same behaviors and inter-model relationships as before. This is coupled with the intuitive understanding that as the gap widens, the Lorentz factor will decrease.} 
    \label{Fig:Lorentz_Flux_Comp}
\end{figure*}

\indent Next, the Lorentz factor of the gap is examined in (Fig. \ref{Fig:Lorentz_Flux_Comp}). Beginning with the unnormalized plots, we see an inverse relationship between increasing mass and decreasing Lorentz factor. For the new model, as the gap reaches the equatorial plane where the gap was smaller, the Lorentz factor slightly increases, as opposed to the exponential decrease seen in the old model. This inverse relationship between Lorentz factors (and by extension $\gamma$-ray energies) and gap width can be attributed to the $E_{||}$ field within the gap. The bigger the gap present, the smaller electric field is needed to accelerate particles to the cascade condition. When the cascade begins, it begins to screen the field out. Viewing the normalized trends, the updated model continues to show a more complex trend as well as bunching near the pole of the KBH. These trends show a counter-intuitive $cos(cos(x))$ dependence that encapsulates the distinct double inflection line see Appendix D. \\

\begin{figure*}
    \centering
    \begin{subfigure}
        \centering
        \includegraphics[scale=0.55]{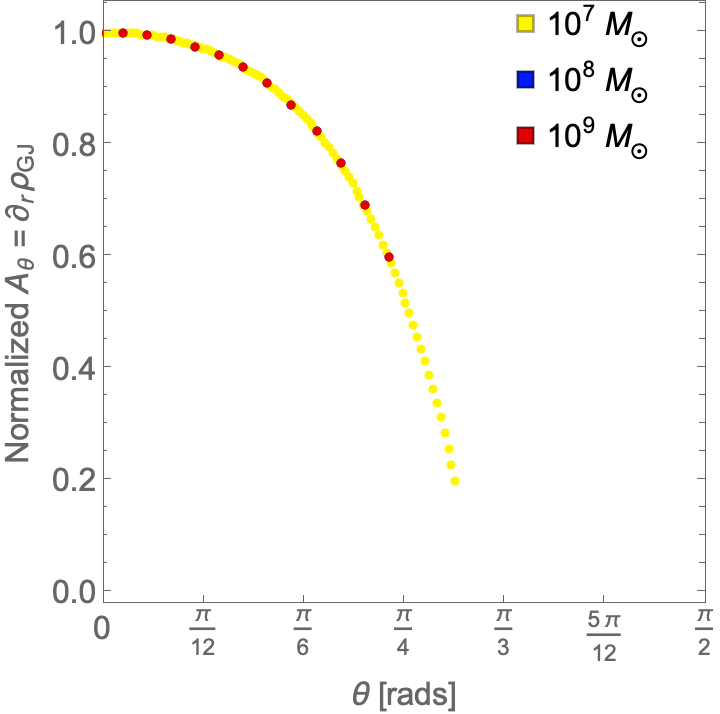}
    \end{subfigure}
    \begin{subfigure}
        \centering 
        \includegraphics[scale=0.55]{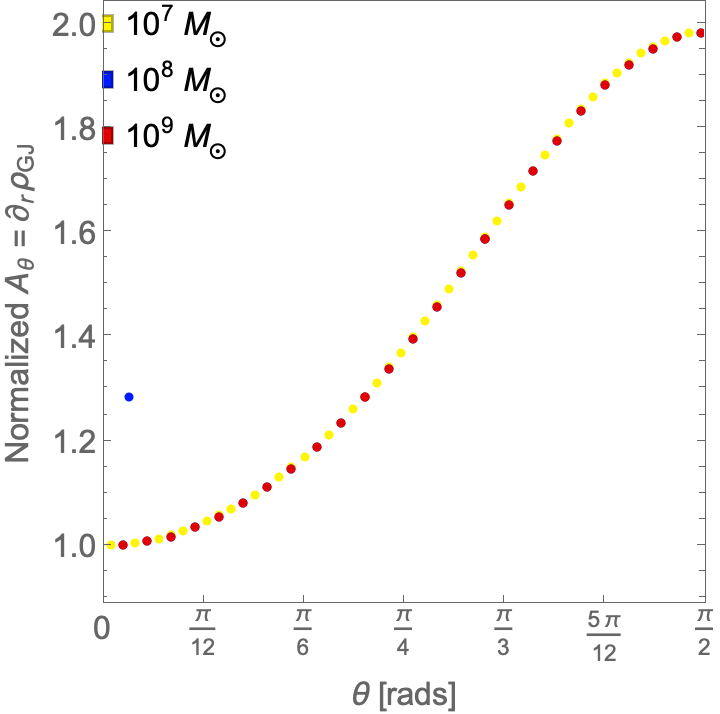}
    \end{subfigure}
    \vskip\baselineskip
    \begin{subfigure}
        \centering 
        \includegraphics[scale=0.55]{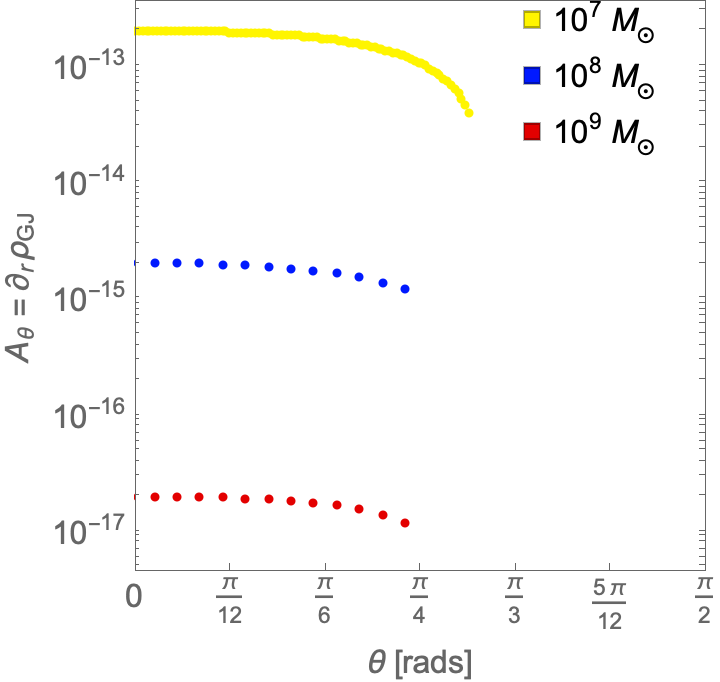}
    \end{subfigure}
    \begin{subfigure}
        \centering 
        \includegraphics[scale=0.55]{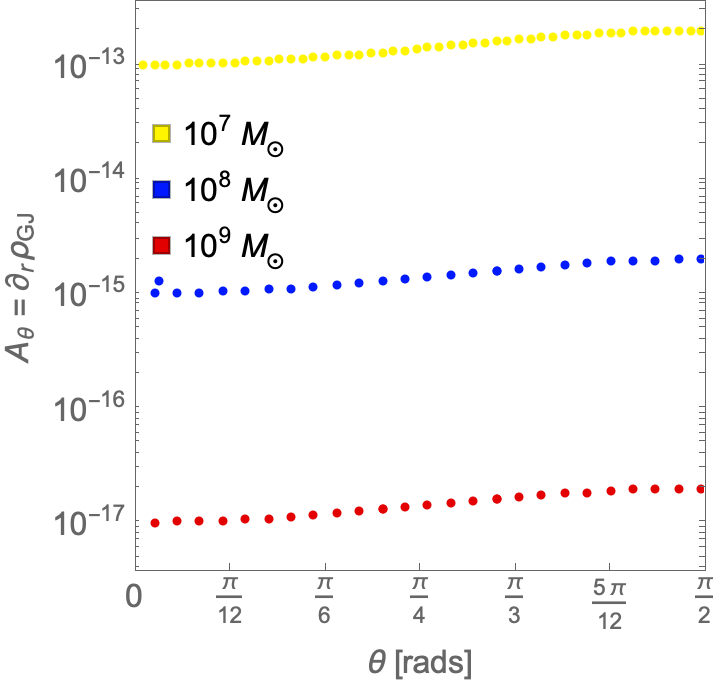}
    \end{subfigure}
    \caption{Comparison of $A_{\theta} = \partial_r\rho_{GJ}$ in old and new models (left and right panels, respectively). The old model ($sin^2(\theta))$ is plotted on the left with the normalized trend above the unnormalized trends. The unnormalized trends are plotted on the logarithmic scale. We see that for both flux models, the mass is inversely related to the trend's magnitude, but does not affect how the trend behaves. In contrast, the flux function does affect the trend behavior.}
    \label{Fig:A_Theta_Flux_Comp}
\end{figure*}

\indent Following the Lorentz Factor, the value $A_{\theta} = \partial_r\rho_{GJ}$ is studied (Fig. \ref{Fig:A_Theta_Flux_Comp}). This factor shows an inverse dependence on the magnitude of the trend, but the behavior of the trend is not affected by the varying mass. In the normalized plots, the new flux continues to show a polynomial type trend that increases as $\theta$ increases around the BH, compared to the decreasing exponential given by the $sin^2(\theta)$ flux model. The unnormalized trends again show a steadily increasing behavior opposite the linear to decreasing trends of the old flux.\\

\begin{figure*}
    \centering
    \begin{subfigure}
        \centering
        \includegraphics[scale=0.6]{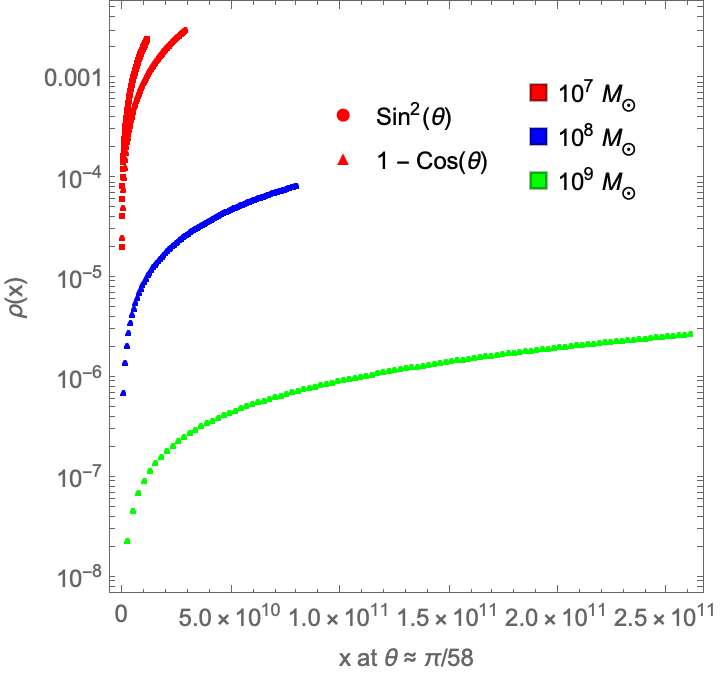}
    \end{subfigure}
    \begin{subfigure}
        \centering 
        \includegraphics[scale=0.6]{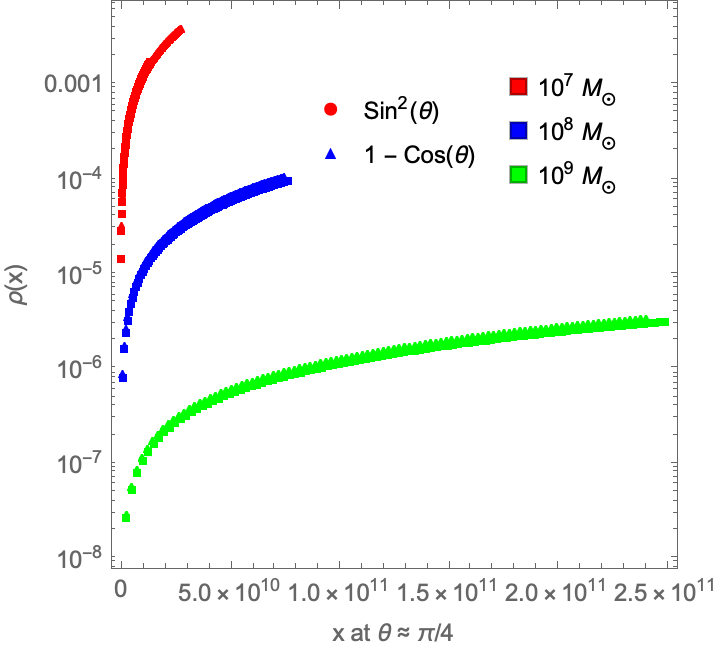}
    \end{subfigure}
    \vskip\baselineskip
    \begin{subfigure}
        \centering 
        \includegraphics[scale=0.6]{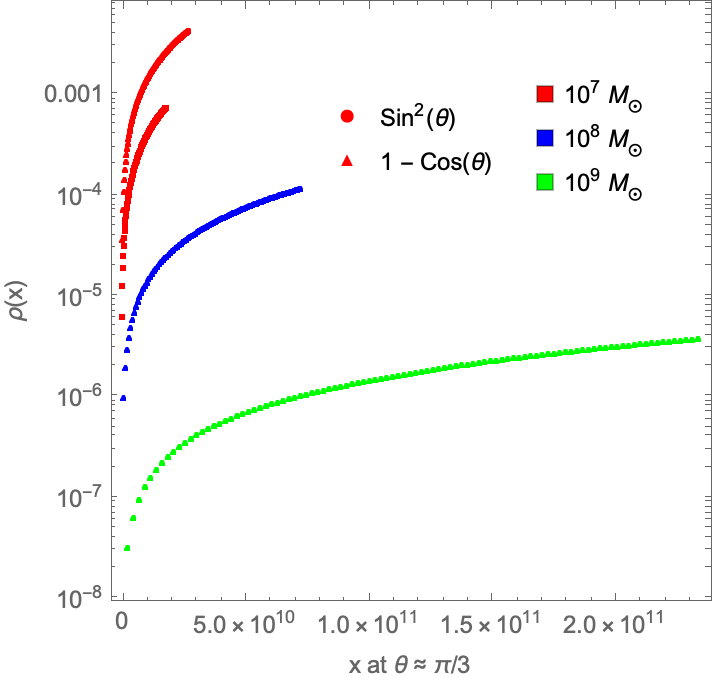}
    \end{subfigure}
    \begin{subfigure}
        \centering 
        \includegraphics[scale=0.6]{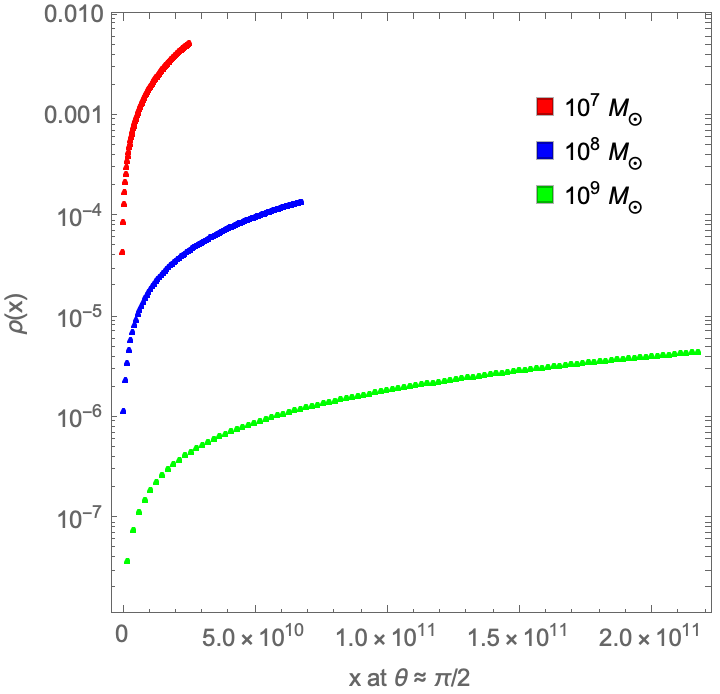}
    \end{subfigure}
    \caption{Comparison of plasma density ($\rho$) as a function of gap coordinate ($x$ in cm}) from the center of the gap at $x = 0$ to the gap edge at $x = H$, where $H$ is the half gap width, for three different BH masses (line colors) for both flux models (shape of point) when applicable. The panels represent different, discrete angle $\theta$ values around the BH where $\theta = 0$ is the axis of rotation and $\theta = \frac{\pi}{2}$ is the equator. The square boxes in the legend represent color, while the shapes on the left of the legend denote which flux is being plotted. By definition, the electric field reach zero at the edge. The difference in gradient of the trends is explained by lower masses having a higher starting field magnitude, thus, they must reach zero in a shorter distance. The divergence seen at low mass can be attributed to a numerical issue due to the stiffness of the system of equations.
    \label{Fig:pvX_Theta_Mass_Comp}
\end{figure*}

\begin{figure*}
    \centering
    \begin{subfigure}
        \centering
        \includegraphics[scale=0.6]{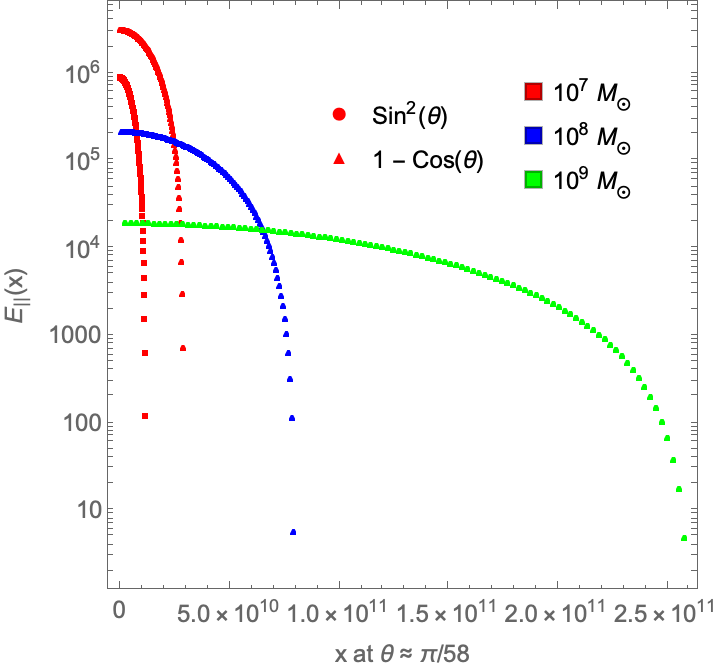}
    \end{subfigure}
    \begin{subfigure}
        \centering 
        \includegraphics[scale=0.6]{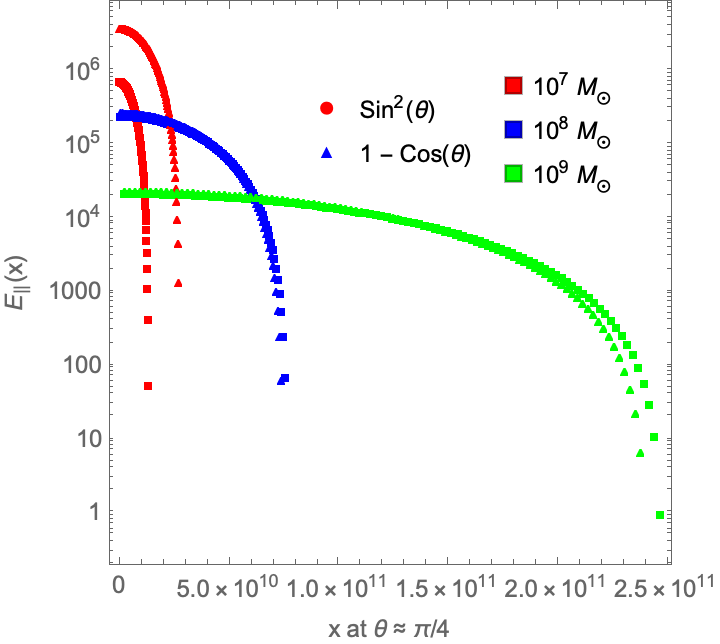}
    \end{subfigure}
    \vskip\baselineskip
    \begin{subfigure}
        \centering 
        \includegraphics[scale=0.6]{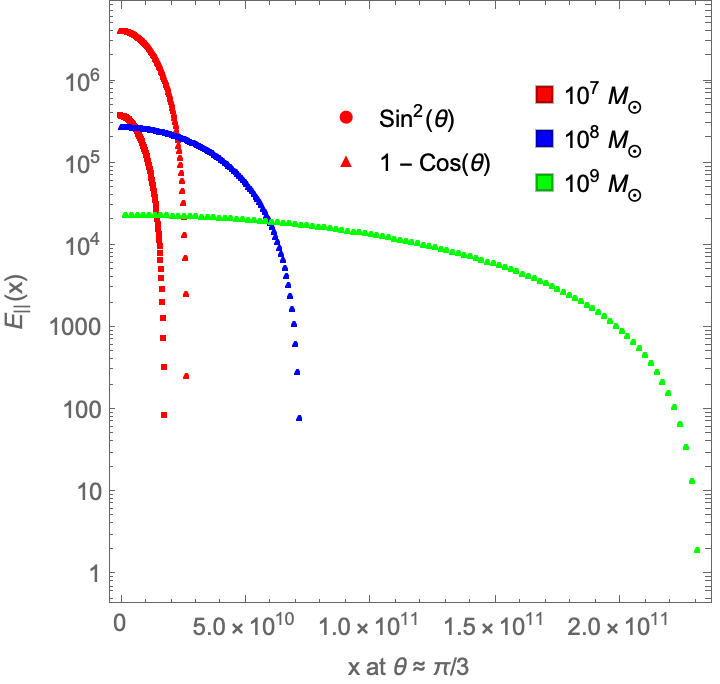}
    \end{subfigure}
    \begin{subfigure}
        \centering 
        \includegraphics[scale=0.6]{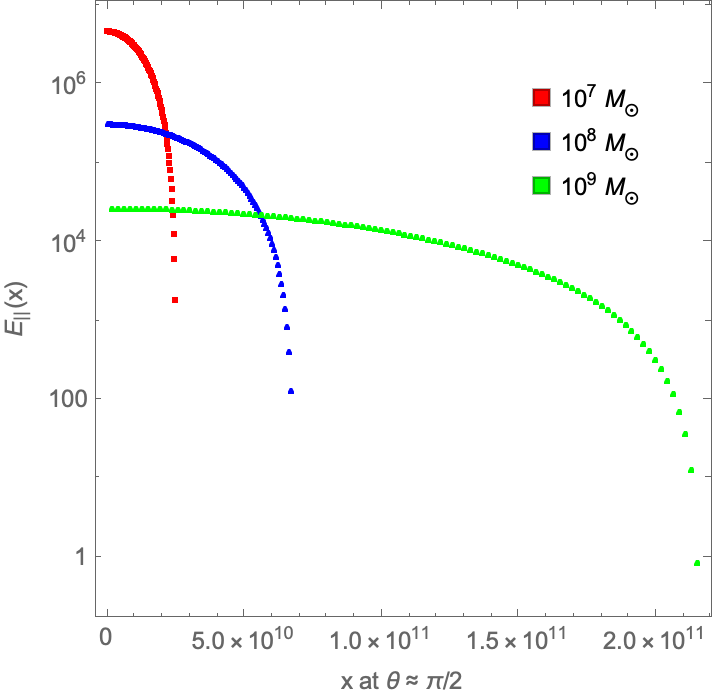}
    \end{subfigure}
    \caption{Comparison of electric field ($E_{||}$) (in statvolt/cm)} as a function of gap coordinate ($x$ in cm) for three different BH masses (line colors) for both flux models (shape of point) when applicable. The square boxes in the legend represent color, while the shapes on the left of the legend denote which flux is being plotted. The panels represent different, discrete angle $\theta$ values around the BH where $\theta = 0$ is the axis of rotation and $\theta = \frac{\pi}{2}$ is the equator.
    \label{Fig:EvX_Theta_Mass_Comp}
\end{figure*}

\begin{figure*}
    \centering
    \begin{subfigure}
        \centering
        \includegraphics[scale=0.6]{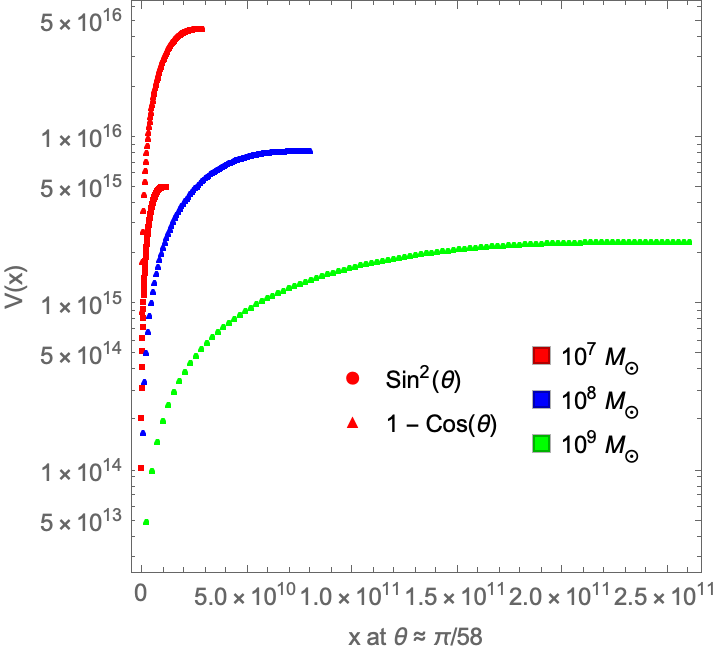}
    \end{subfigure}
    \begin{subfigure}
        \centering 
        \includegraphics[scale=0.6]{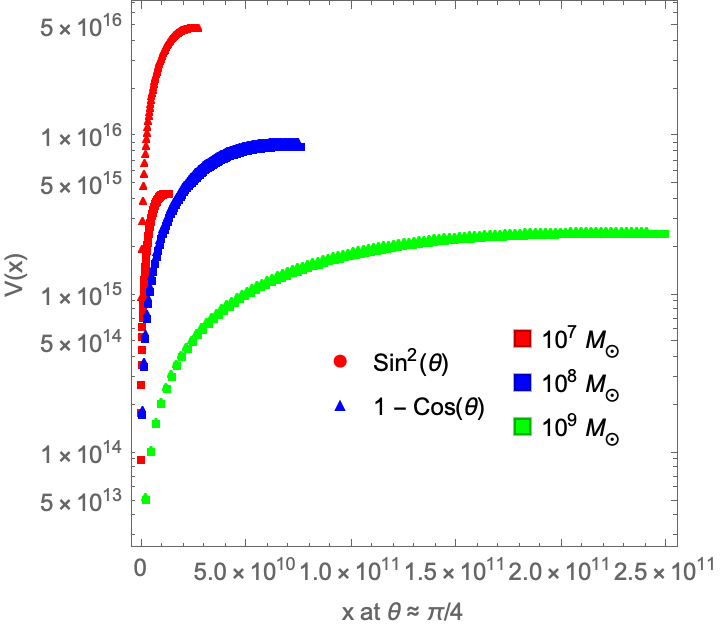}
    \end{subfigure}
    \vskip\baselineskip
    \begin{subfigure}
        \centering 
        \includegraphics[scale=0.6]{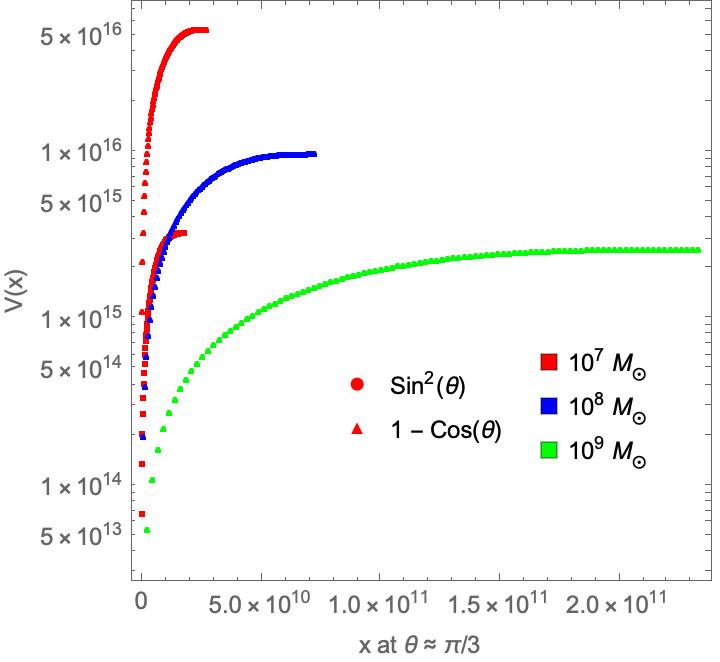}
    \end{subfigure}
    \begin{subfigure}
        \centering 
        \includegraphics[scale=0.6]{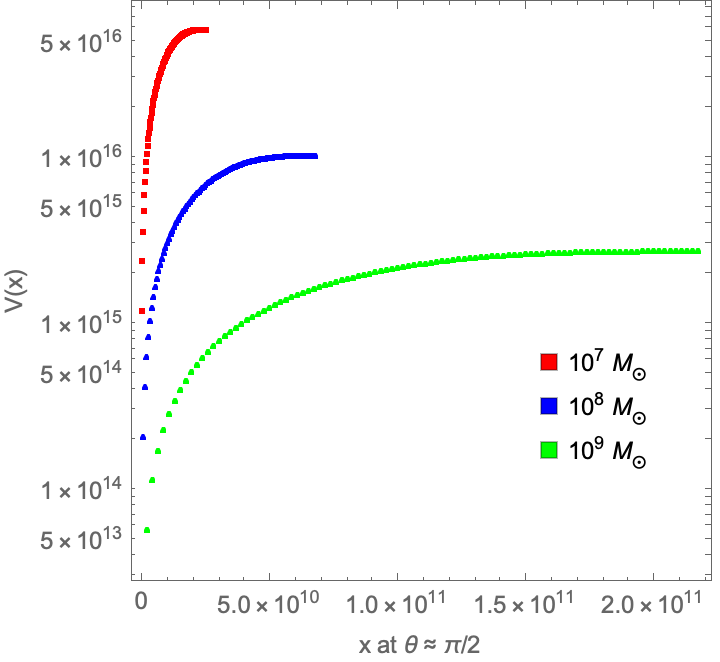}
    \end{subfigure}
    \caption{Comparison of voltage ($V(x)= \int_0^x E_\parallel (x) dx$) (in statvolt) as a function of gap coordinate ($x$ in cm) for three different BH masses (line colors) for both flux models (shape of point) when applicable. The panels represent different, discrete angle $\theta$ values around the BH where $\theta = 0$ is the axis of rotation and $\theta = \frac{\pi}{2}$ is the equator. The square boxes in the legend represent color, while the shapes on the left of the legend denote which flux is being plotted. The $H/2$ voltage peak is another result of the electric field vanishing at the boundaries. With the inverse relation ship between electric field strength and distance, as the Electric field tends to zero, the coordinates continue to rise, giving way to a parabolic fit to the voltage patterns in the gap}
    \label{Fig:VvX_Theta_Mass_Comp}
\end{figure*}

\indent The last three figures (Figs. \ref{Fig:pvX_Theta_Mass_Comp}, \ref{Fig:EvX_Theta_Mass_Comp}, \ref{Fig:VvX_Theta_Mass_Comp}) discussed show trends at discrete polar angles (with the axis of rotation set to $\theta = 0$). These angles are (from closest to axis of rotation to equator): $\pi/58$, $\pi/4$, $\pi/3$, and $\pi/2$. The most striking observation is seen in the length of the trends as the mass increases. This is due to effect of Gauss's law at the  outer boundary position. $x = H$ (outer boundary) is determined by the condition that the Goldreich-Julian charge density $\rho_{GJ}(x = H)$ matches the actual charge density $\rho$, which is given by the current density $J$ divided by $c$. Here, $J$ is conserved along each flow line because of stationality. Because boundary conditions dictate no particle injections at boundaries, the leptons are composed of either electrons or positrons at the boundary, which means that the real charge density $\rho$ is given by $\frac{J}{c}$ at $x = H$. Therefore, if the gap half width $H$ is measured in gravitational radius, $H$ is solely determined as a function of the dimensionless conserved current density, $ j = \frac{J}{\rho_{GJ}(x = H)c}$. Subsequently, if we compare at the same dimensionless current $j$, the actual gap width (e.g., in cm) increases with increasing BH mass, because the gravitational radius increases accordingly. The second overarching observation is the relative angular symmetry within the three trends. While the previously discussed trends had a more direct relation to the $\theta$ dependence of the flux model, the following trends including $E_{||}(x)$, $\rho(x)$, and voltage $V(x)$ appear to show a minimal dependence on $\theta$. Despite a small sample size, the figures show minimal difference in magnitude and overall trend behavior across the entire calculated area for a given plotted value. This is a primary observation, as the entire calculated range ($0 - \frac{\pi}{2}$) is only represented by $4$ polar angles. Finally, the difference in trends between the flux models. For the first three $\theta$ values in (Figs. \ref{Fig:pvX_Theta_Mass_Comp}, \ref{Fig:EvX_Theta_Mass_Comp}, \ref{Fig:VvX_Theta_Mass_Comp}), there are two trend lines per mass, where the square data point represents the old flux model and the triangle data point represents the masses in the new flux model. For each $\theta$ value where both fluxes are plotted, the difference between the two fluxes for $10^8 M_{\odot}$ and $10^9 M_{\odot}$ is noticeable, but minimal relative to the difference generated by the difference in mass. This includes the previously discussed observations (trend length along the x-axis and angular symmetry) The mass containing the largest differential between fluxes is the lowest mass. However, the trend lines for $10^7 M_{\odot}$ is consistently approximately one magnitude apart.\\

\indent The first trend (Fig. \ref{Fig:pvX_Theta_Mass_Comp}) depicts the plasma density ($\rho$) as a function of $x$ in cm from the center of the gap at $x = 0$ to the gap edge at $x = H$, where $H$ is the half gap width. Along with the observations previously discussed, the trends show a direct relation between mass of the BH and the steepness of the rise in plasma density from center to edge of the gap. This difference in slope can be seen at two points on the trend. As the lines begin, there is a steep increase in plasma density right after $x = 0$, after which is an area of inflection where the trend levels out and continues the remaining way to the edge of the gap, $x = H$. It can be seen that the higher the mass of the BH, the lower these two slopes are in relation to lower BH masses. Recall, that the boundary of the gap $x = H$ is defined as the region where plasma density $\rho$ is equal to $\rho_{GJ}$. The null surface $x = 0$ is defined as the region $\rho_{GJ} = 0$. As the charge density ($\rho_{GJ}$) increases (rapidly from $0$ at first, then gradually as $x$ increases), plasma density rises as well until the two meet. The steepness in trends can be explained by the relative distance the plasma density has before it meets the charge density. With a thinner gap, the density gradient must be larger to cover the same two magnitudes the higher mass systems who posses a smaller gradient over a much larger distance.\\

\indent Next, the electric field $E_{||}$ (in statvolt/cm) (Fig. \ref{Fig:EvX_Theta_Mass_Comp}) as a function of gap coordinate center (in cm) $x = 0$ to edge $x = H$ is plotted. The electric field peaks, for each mass, just after the null surface and exponentially drops as it approaches the edge. This is a result of the electric field being screened as it approaches the edge of the gap. Also recall, that by definition, the electric field must be zero at the edge. This explains each mass tending to zero as $x$ approaches $H$. The difference in gradient of the masses is explained by this concept. Because the lower masses have higher starting field magnitude, they must reach zero in a shorter distance. \\

\indent This can be shown if we recall
\begin{equation}\label{Eq:dRhodX_From_Ref4}
    \frac{d\rho_{GJ}}{dx} \approx \frac{B(\Omega - \omega)}{H},
\end{equation}
and
\begin{equation}\label{Eq:EPar_From_Ref4}
    E_{\parallel}(x) \propto \frac{d\rho_{GJ}}{dx}(H^2 - x^2),
\end{equation}
\citep{21}, where $\rho_{GJ}$ denotes the Goldreich-Julian charge density, $\Omega$ denotes the angular frequency of  the magnetic field lines, $\omega$ does the frame-dragging angular frequency at position $x$ in the gap, $B$ the magnetic field strength at point $x$, and  $ r_{g}= GM/c^2 $ the gravitational radius. After some algebra, (Eqs. \ref{Eq:dRhodX_From_Ref4} and \ref{Eq:EPar_From_Ref4}) can be related through
\begin{equation}\label{Eq:EP_Propto_Ref4}
    E_\parallel \propto \frac{B}{r_g} \propto \frac{M^{-1/4}}{M} = M^{-5/4}.
\end{equation}
Note that dimensionless gap width $H/r_{g}$ cancels out in the RHS.
Thus, $E_\parallel \propto M^{-5/4}$, which is shown by the curves in (Fig. \ref{Fig:EvX_Theta_Mass_Comp}). As $E_{||}(x)$ (cgs) increases, so does $\Gamma(x)$ for $e^-$'s. Which in turn creates in an increased pair production rate, resulting in more efficient screening of $E_{||}(x)$ near the outer edge at $x = H$. This results in a smaller gap $H$ for lower BH masses $M$, this confirming the previous statement.\\

\indent Finally, the voltage drop in statvolt between the center of the gap (in cm) and a point in the gap is plotted and studied \ref{Fig:VvX_Theta_Mass_Comp}. The voltage in the gap is calculated via
\begin{equation}
   V(x)= \int_0^x E_\parallel (x) dx
\end{equation}
where $dx$ is the x-coordinate (the center of the gap is $x = 0$) interval. Again, some symmetries and relations can be immediately deduced. The relative difference between the old and new flux modes (with the one exception) and the difference between the different angles plotted both continue to be minimal. We see that as $E_{||}$ decreases as $x$ increases in (Fig. \ref{Fig:EvX_Theta_Mass_Comp}), the voltage's gradient correspondingly remains positive throughout the gap. The voltage increases with a steepness corresponding to mass (and therefore gap width) with steeper trends being seen at lower mass (thinner gap width). \\

\indent Looking at all of the trends from the $(1 - cos(\theta))$ model, we examine the relationships between the three main parameters within the model. The trends for $A_{\theta}$ and voltage shows some changes based on flux, but mainly the difference can be explained by the previous values or the varying mass. Despite the wildly different functional expressions for the trends, we can derive behaviors in the gap seen across the three masses. We can see a direct relationship between the Lorentz factor and the outgoing flux from the gap. As the ZAMO moves down from the pole to the equator, more radiation is produced. \\

\clearpage
\newpage

\subsection{Changing of Spectral Index, $\alpha$}
\begin{figure}
    \centering
    \begin{subfigure}
        \centering
        \includegraphics[scale=0.5]{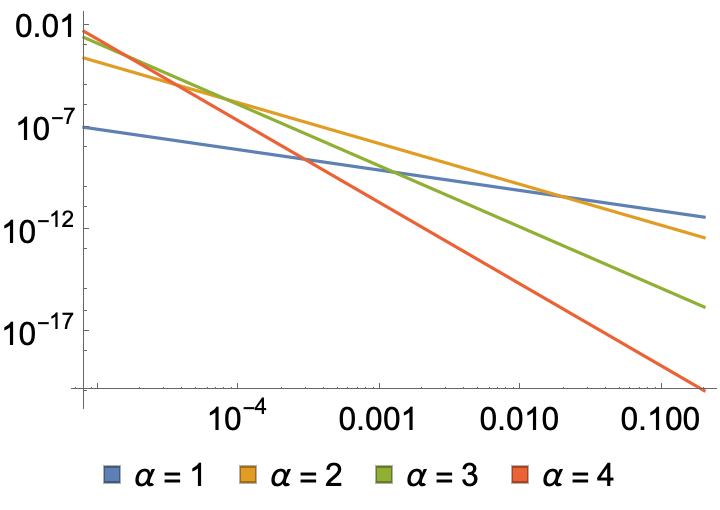}
    \end{subfigure}
    \hfill
    \begin{subfigure}
        \centering 
        \includegraphics[scale=0.5]{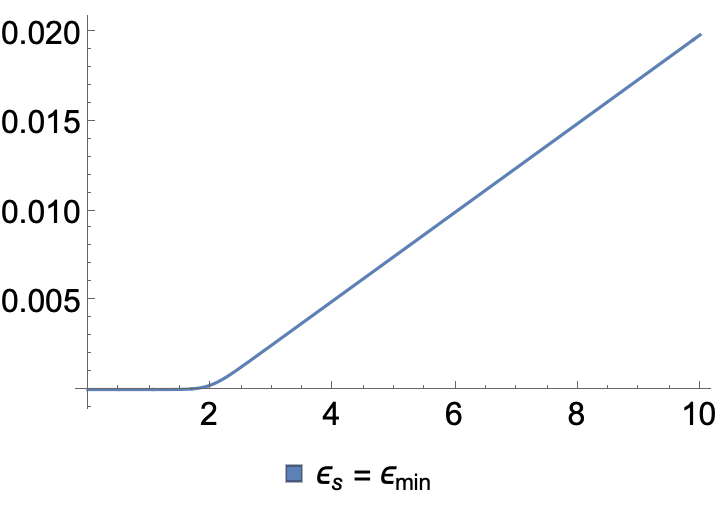}
    \end{subfigure}
    \vskip\baselineskip
    \begin{subfigure}
        \centering 
        \includegraphics[scale=0.5]{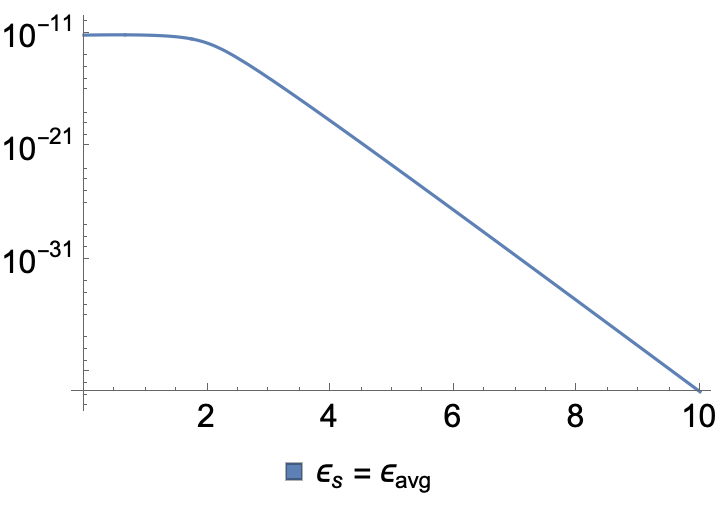}
    \end{subfigure}
    \hfill
    \begin{subfigure}
        \centering 
        \includegraphics[scale=0.5]{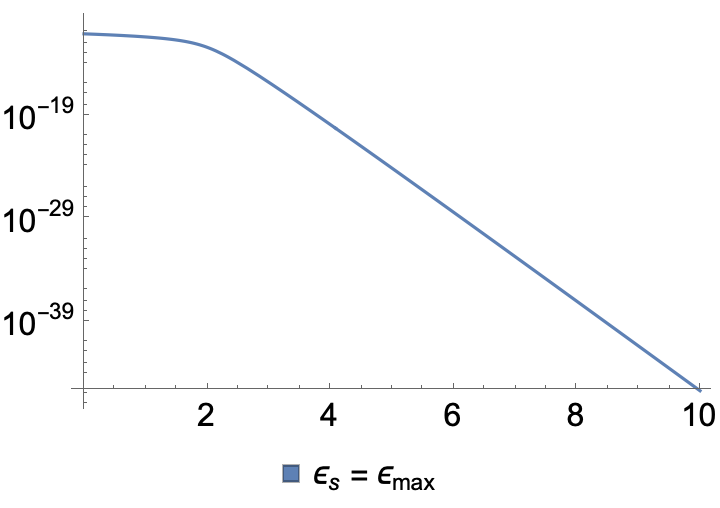}
    \end{subfigure}
    \caption{In the top right, the spectral number density as a function of photon energy is plotted for different levels of spectral index ($\alpha$) in log-log scale. The functional trend as a function of energy across the differing indices illuminates the challenges of sweeping the system of equations. The remaining three plots show the spectral number density as a function of alpha for the minimum, average (semi-log), and maximum (semi-log) photon energy used in this study. The function's most drastic changes all center around the initial condition of $\alpha = 2$, creating a challenge to sweep to other spectral indices.} 
    \label{Fig:Alpha_Energy_Predictions}
\end{figure}

\indent The first challenges to the extended spectral study of $e^\pm$ cascade in the magnetosphere of black holes comes with changing the spectral number density of ambient soft photons of the system of equations.  As expressed in an earlier section, the system of equations that governs this system is rigid, and pushes back any changes to the equations when the system is stable. Viewing the spectral number density as a function of photon energy for different $\alpha$ in (Fig. \ref{Fig:Alpha_Energy_Predictions}) illuminates the drastic difference of function between integer spectral index numbers. The remaining panels show the spectral energy density as a function of $\alpha$ for the minimum, average, and maximum input spectral energy. The steep curves and sharp inflection is the source of the challenge when it comes to sweeping the index variable. The abrupt change in not ideal for the shooting method and this system of equations, causing the calculations require immense computational resources.\\ 

\begin{figure}
    \centering
    \begin{subfigure}
        \centering
        \includegraphics[scale=0.5]{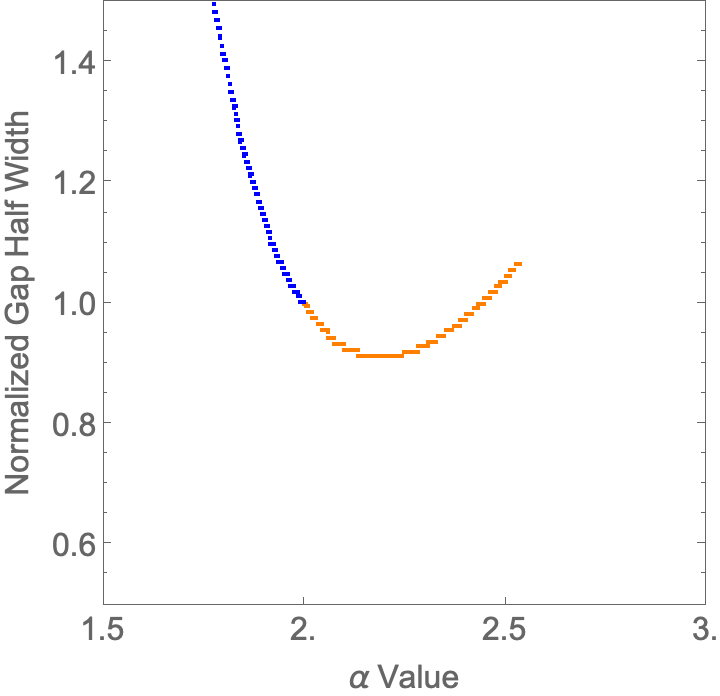}
    \end{subfigure}
    \begin{subfigure}
        \centering 
        \includegraphics[scale=0.5]{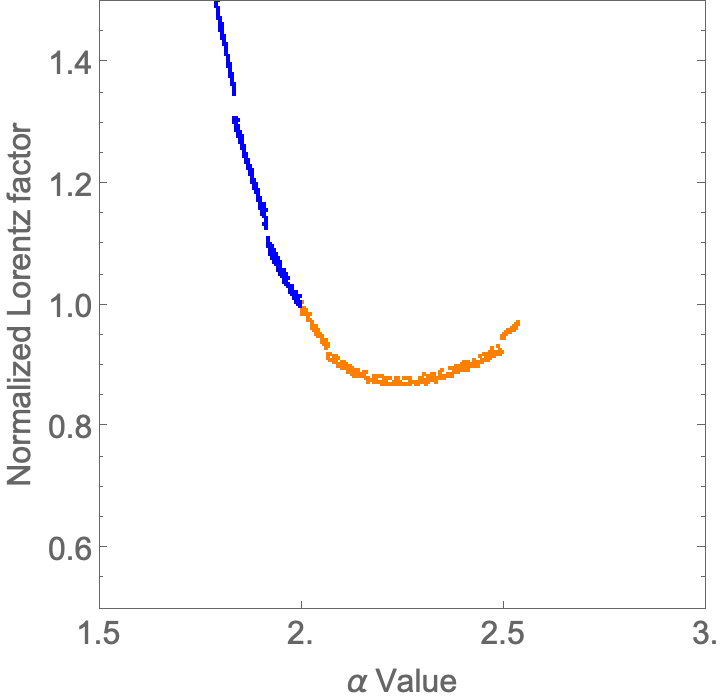}
    \end{subfigure}
    \vskip\baselineskip
    \begin{subfigure}
        \centering 
        \includegraphics[scale=0.5]{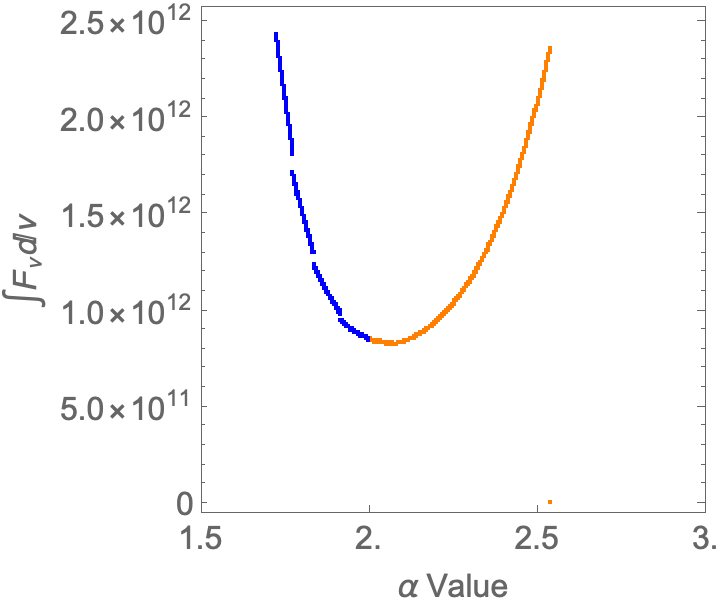}
    \end{subfigure}
    \caption{Depictions of the three main parameter trends (half width, Lorentz factor, and outgoing flux) as a function of $\alpha$. The different colors represent different data sets from the simulation as spectral index is swept from $\alpha = 2$. The sharp inflection seen around $\alpha = 2$ is believed to be the main challenge.} 
    \label{Fig:Alpha_Change_3_Values}
\end{figure}

\indent While changing the spectral index has stretched computational resources and the analysis of different spectral states incident on the KBH system is ongoing, the main trends as a function of $\alpha$ at the axis of rotation have been recovered (Fig. \ref{Fig:Alpha_Change_3_Values}). While half gap width and Lorentz factor show a manageable change in parameter, the outgoing flux has posed certain numerical challenges.  

\indent The change in trend of the half gap width, Lorentz factor, and particle flux as a function of spectral index illustrates the drastic inflections encountered when the code is solving the BVP, shown in (Fig. \ref{Fig:Alpha_Change_3_Values}). These inflections are the main challenge when solving for different spectral states, as the code must begin at the known solution ($\alpha = 2$) to solve for the next spectral state. The different colors of the line signify the various data sets used to create a semi-continuous trend. The code runs for such a extended period of time that a single sweep is not feasible at this time. We emphasize that the $\alpha = 2$ spectrum yields the most favorable conditions for the pair cascade. This is seen from the minimum Lorentz factor and minimum gap width. Softer ($\alpha > 2$) and harder spectra ($\alpha < 2$) reduce the cascade efficiency.

\section{Results and Discussion}
\indent In this paper we explored the role of the magnetic field configuration around a spinning BH on the electron-positron cascade and its properties in particular. Two models, representing a thin accretion disk and a quasi-spherical gas distribution were compared. We have shown that the new magnetic flux model for a KBH $e^\pm$ cascade has not only shown the expected oppositely directed trends, but delivered a more complex mathematical relationship as a function of polar angle. In comparison with the original flux model of $sin^2(\theta)$, $(1 - cos(\theta))$ allows a more realistic treatment of accretion disks far way from the KBH, close disks that are filled with thick, hot material, and systems that have no disk at all such as the Milky Way.  \\

\indent This paper details the beginning of an extended study on this new flux model as well as the first computational challenges seen when beginning a study on the spectral states of KBH systems. This study suggests that the parameter trends as a function of $\theta$ are not simple monotonic functions, but more complex expressions that have a greater variance withing the differing parameter sets, such as the varying mass shown in this work. The intuitive nature of behaviors in the gap remain; the $\gamma$-ray photon flux correlates with the max Lorentz factor and anti-correlates with the gap width. \\

\indent Particularly interesting results of this study concern the efficiency of the pair cascade. First, the efficiency of the pair cascade in a quasi-spherical gas distribution case studied here increases with latitude and peaks at the jet axis. In contrast, in the thin disk accretion case, highest efficiency shown around the jet-disk interface. This finding is prediction of our study and a possible observational diagnostic of the field geometry at astrophysical sources. Second, the the spectrum of background photons is also a key factor in the leptonic cascade phenomenon. The efficiency is maximum for a spectra index of two. Both harder and softer spectra result is a less efficient leptonic cascade.\\

\indent Follow up studies are underway examining different parameters and will be readily compared to the results found in \citep{20}. The physical and spectral parameter space available can help shed light on how these systems utilize the BZ mechanism to fuel powerful general relativistic jets seen in observations.

\begin{acknowledgments}
This work was supported by NSF grant PHY-2010109.
\end{acknowledgments}

%






\clearpage
\newpage
\appendix

\section{Full Line Element Expression}
\begin{multline}\label{Eq:Full_Line_Element}
    ds^2 = -\left(\frac{(r^2 - (2GM/c^2)r + (J/Mc)^2) - a^2sin(\theta)}{r^2 + (J/Mc)^2cos^2(\theta)}\right)c^2dt^2 
    - \left(\frac{2(2GM/c^2)(J/Mc)sin^2(\theta)c}{r^2 + (J/Mc)^2cos(\theta)}\right)dtd\psi \\
    + \left(\frac{((r^2 + (J/Mc)^2) - (r^2 (2GM/c^2)r + (J/Mc)^2)(J/Mc)^2sin^2(\theta))sin^2(\theta)}{r^2 + (J/Mc)^2sin(\theta)}\right)d\psi^2
    \\ + \left(\frac{r^2 + (J/Mc)^2cos^2(\theta)}{r^2 - (2GM/c^2)r + (J/Mc)^2}\right)dr^2 + (r^2 + (J/Mc)^2cos^2(\theta))d\theta^2.
\end{multline}

\clearpage
\newpage

\section{Initial Conditions}
\begin{longtable}{ |p{7cm}||p{5cm}|p{5cm}|} 
\caption{The initial conditions (ICs) for the old flux model ($sin(\theta)^2$) and new flux model ($(1 - cos(\theta))$.}
\label{Tab:Initial_Conditions}\\
 \hline
 Simulation Code Parameter & $sin(\theta)^2$ ICs & $(1 - cos(\theta))$ ICs \\
 \hline
 BH Mass (Comp. Units) & $2\times10^{40}$ & $2\times10^{40}$\\
 BH Mass (Physical Units) & $1 \times 10^7 M_{\odot}$ & $1 \times 10^7 M_{\odot}$ \\
 BH Spin Factor & $0.999999$ & $0.999999$ \\
 Half Gap Width & $25516783911.8834$ & $25516783911.8834$ \\
 Lorentz Factor at Center ($\Gamma_0$) & $14195.3939198398$ & $14195.3939198398$ \\
 Lorentz Factor at Edge ($\Gamma_H$) & $2219.643845290660465252586$ & $2219.643845290660465252586$ \\
 Number of Energy Bins & $51$ & $51$\\
 Bin Spacing & $\sqrt{2}$ & $\sqrt{2}$ \\
 $\epsilon_{Max}$ ($\frac{Energy}{m_ec^2}$) & $2.0\times10^{-1}$ & $2.0\times10^{-1}$\\
 $\epsilon_{Min}$ ($\frac{Energy}{m_ec^2}$) & $8.0\times10^{-6}$ & $8.0\times10^{-6}$\\
 $\rho_{GJ}$ & $5.0\times10^{-16}$ & $5.0\times10^{-16}$ \\
 $j_0 = AH\sqrt{1.0-\frac{1}{\Gamma_H^2}}$ & $0.0000127584$ & $0.0000127584$ \\
 Field Drag Parameter & $fd=0.45$ & $fd=0.45$ \\
 $B_{||}$ (Initial $B$ Field) & $1.0\times10^4$ & $1.0\times10^4$ \\
 $\Psi_0$ (Initial Flux) & $1.0\times10^{32}$ & $1.0\times10^{32}$\\
 Background Field Energy Density ($\frac{U_b\sigma_T}{m_ec^2}$) & $0.8125532\times10^{-13}$ & $0.8125532\times10^{-13}$\\
 \hline
\end{longtable}

\section{Boundary Conditions Table}

\begin{longtable}{ |p{4cm}|p{5.5cm}|p{6cm}|p{1.5cm}|} 
\caption{The boundary conditions for the SMBH system cascade equations in the gap, recreated from \citep{20}}
\label{Tab:Boundary_Condtions}\\
 \hline
Condition & Equation Used & Symmetric Assumption & Boundary \\
 \hline
 $E_{||} = (\Gamma^2 - 1)\sigma_TU_be^{-1}$ & $\frac{m_ec^2d\Gamma}{dx} = eE_{||} - (\Gamma^2 - 1)\sigma_TU_b$ & $E_{||}(x) = E_{||}(-x)$ and $\Gamma(x) = \Gamma(-x)$ & $x=0$ \\
 $2n^+\sqrt{1-\frac{1}{\Gamma^2}} = \frac{j_0}{e}$ & $j_0 = e[n^+(x) + n^-(x)]\sqrt{1 - \frac{1}{\Gamma^2(x)}}$ & $n^+(x) = n^-(-x)$ & $x=0$ \\
 $f^+_i = f^-_i$ & $f^\pm_i(x) \equiv \int_{\xi_{i-1}}^{\xi_i}F^\pm(x,\epsilon_\gamma)d\epsilon_\gamma$ & $F^+(x) = F^-(-x)$ & $x=0$ \\
 $E_{||} = 0$ & $\frac{dE_{||}}{dx} = 4\pi[e(n^+ - n^-) - \rho_{GJ}$  & $\rho_{gap} = \rho_{GJ}$ & $x=H$ \\
 $n + \sqrt{1 - \frac{1}{\Gamma^2}} = \frac{j_0}{e}$ & $j_0 = e[n^+(x) + n^-(x)]\sqrt{1 - \frac{1}{\Gamma^2(x)}}$ & $n^-(x) = 0$ & $x=H$ \\
 $j_0(1 - \frac{1}{\Gamma^2})^{-1/2} -Ax = 0$ & $j_0 = e[n^+(x) + n^-(x)]\sqrt{1 - \frac{1}{\Gamma^2(x)}}$ & $\frac{dE_{||}}{dx} = 0$ & $x=H$ \\
 $f^-_i = 0$ & $f^-_i(x) \equiv \int_{\xi_{i-1}}^{\xi_i}F^-(x,\epsilon_\gamma)d\epsilon_\gamma$ & $F^-(x) = 0$ & $x=H$ \\
 \hline
\end{longtable}

\clearpage
\newpage

\section{Approximate Trend Fits}

\begin{center}
\begin{longtable}{|p{9cm}|p{8cm}|}
\caption{The trend fits for the plots shown in the body of the paper.} \label{Tab:Trend_Fits} \\

\hline \multicolumn{1}{|c|}{\textbf{Figure and Line}} & \multicolumn{1}{c|}{\textbf{Functional Fit}} \\ \hline 
\endfirsthead

\multicolumn{2}{c}%
{{\bfseries \tablename\ \thetable{} -- continued from previous page}} \\
\hline \multicolumn{1}{|c|}{\textbf{Figure and Line}} &
\multicolumn{1}{c|}{\textbf{Functional Fit}}  \\ \hline 
\endhead

\hline \multicolumn{2}{|c|}{{Continued on next page}} \\ \hline
\endfoot

\hline \hline
\endlastfoot

 \hline
 Change of Flux: Norm. Gap Half Width & $1 + 0.019e^{2.141 x}$ \\ 
 Change of Flux: Norm. Lorentz Factor & $1 - 0.032e^{1.956 x}$ \\
 Change of Flux: Norm. Outgoing Flux & $1 - 0.016e^{4.281 x}$ \\
 \hline
 \hline
 Flux Comparison: $sin^2(\theta)$ Norm. Gap Half Width & $1 + 0.75x^5$ \\
 Flux Comparison: $sin^2(\theta)$ Norm. Lorentz Factor & $1 - 0.03x - 0.08x^2$ \\
 Flux Comparison: $sin^2(\theta)$ Norm. Outgoing Flux & $0.1 + 0.9cos(x^2)$ \\
 Flux Comparison: $sin^2(\theta)$ Norm. A & $1 - 0.2x^2 - 1.9x^3 + 14.5x^4 -  78.6x^5 + 264x^6 - 577x^7 + 814.9x^8 + 356.6x^{10} - 76.9x^{11} - 716.7|x|^9$ \\
 Flux Comparison: $sin^2(\theta)$ Norm. $j_0$ & $1 - 0.92^{7.7sin(3.2x)}x^3$ \\
 Flux Comparison: $(1 - cos(\theta))$ Norm. Gap Half Width & $1 + 0.03x + 0.1x^2$ \\
 Flux Comparison: $(1 - cos(\theta))$ Norm. Lorentz Factor & $1 - x^3$ \\
 Flux Comparison: $(1 - cos(\theta))$ Norm. Outgoing Flux & $1 + 0.65x^2$ \\
 Flux Comparison: $(1 - cos(\theta))$ Norm. A & $1 + 0.6x^2$ \\
 Flux Comparison: $(1 - cos(\theta))$ Norm. $j_0$ & $2 - cos(x)$ \\
 \hline
 Flux Comparison: $sin^2(\theta)$ Gap Half Width & $1928.85 - 832.465x^4$ \\
 Flux Comparison: $sin^2(\theta)$ Lorentz Factor & $1.188 \times 10^{10} + 7.715 \times 10^9 x^5$ \\
 Flux Comparison: $sin^2(\theta)$ Outgoing Flux & $1.733 \times 10^{12}x$ \\
 Flux Comparison: $sin^2(\theta)$ A & Too Small To Approximate \\
 Flux Comparison: $sin^2(\theta)$ $j_0$ & $0.002 - 0.002x^{2.7 + x}$ \\
 Flux Comparison: $(1 - cos(\theta))$ Gap Half Width & $11423.3 + 373.036x + 1172.47x^2$ \\
 Flux Comparison: $(1 - cos(\theta))$ Lorentz Factor & $2.9 \times 10^{10} - 7.9 \times 10^8 x - 2.27 \times 10^9x^2$ \\
 Flux Comparison: $(1 - cos(\theta))$ Outgoing Flux & $8.5 \times 10^{11} + 5.6 \times 10^{11}x^2 + 3.2 \times 10^7 tan(1.75 x)$ \\
 Flux Comparison: $(1 - cos(\theta))$ A & Too Small To Approximate \\
 Flux Comparison: $(1 - cos(\theta))$ $j_0$ & $0.002 + 0.002x^2$ \\
 \hline
 \hline
 Mass Comparison: $sin^2(\theta)$ Norm. Gap Half Width & $1 + 0.0005e^{7.421x}$ \\
 Mass Comparison: $sin^2(\theta)$ Norm. Lorentz Factor & $1 – 0.0024e^{5.357x}$\\
 Mass Comparison: $sin^2(\theta)$ Norm. Outgoing Flux & $1 – 0.0168e^{4.223x}$ \\
 Mass Comparison: $(1 - cos(\theta))$ Norm. Gap Half Width & $1.013 - 0.109x$ \\
 Mass Comparison: $(1 - cos(\theta))$ Norm. Lorentz Factor & $0.751 + 0.457cos(cos(x))$ \\
 Mass Comparison: $(1 - cos(\theta))$ Norm. Outgoing Flux & $1+ sin(0.63x^2)$\\
 \hline
 Mass Comparison ($10^6 M_{\odot}$): $sin^2(\theta)$ Gap Half Width & $4.1\times10^9 + 446408e^{9x}$ \\
 Mass Comparison ($10^7 M_{\odot}$): $sin^2(\theta)$ Gap Half Width & $1.2\times10^{10} + 7.1\times10^6e^{7.2x}$ \\
 Mass Comparison ($10^8 M_{\odot}$): $sin^2(\theta)$ Gap Half Width & $4.2\times10^{10} + 1.3\times10^7e^{8.1x}$ \\
 Mass Comparison ($10^6 M_{\odot}$): $sin^2(\theta)$ Lorentz Factor & $14205.4 - 39.6e^{5.4x}$ \\
 Mass Comparison ($10^7 M_{\odot}$): $sin^2(\theta)$ Lorentz Factor & $1946.4 - 6.3e^{5x}$ \\
 Mass Comparison ($10^8 M_{\odot}$): $sin^2(\theta)$ Lorentz Factor & $611.8 - 1.6e^{-54837.4x}$ \\
 Mass Comparison ($10^6 M_{\odot}$): $sin^2(\theta)$ Outgoing Flux & $1.5\times10^{13} - 4.5\times10^{11}e^{^3.6x}$ \\
 Mass Comparison ($10^7 M_{\odot}$): $sin^2(\theta)$ Outgoing Flux & $1.4\times10^{11} - 2.8\times10^9e^{4x}$ \\
 Mass Comparison ($10^8 M_{\odot}$): $sin^2(\theta)$ Outgoing Flux & $1.8\times10^9 - 4.4\times10^7e^{3.8x}$ \\
 \hline
 Mass Comparison ($10^7 M_{\odot}$): $(1 - cos(\theta))$ Gap Half Width & $3\times10^{10} - 2.8\times10^9x$ \\
 Mass Comparison ($10^8 M_{\odot}$): $(1 - cos(\theta))$ Gap Half Width & $8.1\times10^{10} - 9.1\times10^9x$ \\
 Mass Comparison ($10^9 M_{\odot}$): $(1 - cos(\theta))$ Gap Half Width & $2.7\times10^{11} - 3.4\times10^{10}x$ \\
 Mass Comparison ($10^7 M_{\odot}$): $(1 - cos(\theta))$ Lorentz Factor & $11044.3 + 2091.8x + 285.6cos(5.3x)$ \\
 Mass Comparison ($10^8 M_{\odot}$): $(1 - cos(\theta))$ Lorentz Factor & $2931 + 477.5x$ \\
 Mass Comparison ($10^9 M_{\odot}$): $(1 - cos(\theta))$ Lorentz Factor & $1064.2 + 164.2cos(x)sin[123.993+x]$ \\
 Mass Comparison ($10^7 M_{\odot}$): $(1 - cos(\theta))$ Outgoing Flux & $8.5\times10^{11} + 8.3\times10^{10}x + 9.9\times10^{11}x^3 - 4.9\times10^{11}x^4$ \\
 Mass Comparison ($10^8 M_{\odot}$): $(1 - cos(\theta))$ Outgoing Flux & $5.8\times10^9 + 5\times10^9x$ \\
 Mass Comparison ($10^9 M_{\odot}$): $(1 - cos(\theta))$ Outgoing Flux & $7\times10^7 + 5.2\times10^7x$ \\
 \hline
 \hline
 Change of Alpha: Norm. Gap Half Width & $518.3 - 1126.3x + 986.8x^2 - 434.3x^3 + 95.9x^4 - 8.5x^5$ \\ 
 Change of Alpha: Norm. Lorentz Factor & $211.76 - 366.1x + 239.3x^2 - 69.8x^3 + 7.7x^4$ \\
 Change of Alpha: Norm. Outgoing Flux & $-48807.5 + 141702x - 170545x^2 + 108927x^3 - 38943.8x^4$ \\
\end{longtable}
\end{center}
\clearpage
\newpage

\bibliography{BH1_Bib}{}
\bibliographystyle{aasjournal}



\end{document}